\documentclass[lettersize,journal]{IEEEtran}

\usepackage[style=ieee, sorting=none, backend=biber]{biblatex}
\usepackage{amsmath,amssymb,amsfonts,amsthm}

\usepackage{graphicx}
\usepackage{textcomp}
\usepackage{xcolor}
\def\BibTeX{{\rm B\kern-.05em{\sc i\kern-.025em b}\kern-.08em
    T\kern-.1667em\lower.7ex\hbox{E}\kern-.125emX}}

\usepackage{url}

\usepackage{booktabs}       %
\usepackage{array}          %
\usepackage{colortbl}       %
\usepackage{stfloats}       %
\usepackage[flushleft]{threeparttable}
\usepackage{amsfonts}       %
\usepackage{nicefrac}       %
\usepackage{microtype}      %
\usepackage{lipsum}
\usepackage{fancyhdr}       %
\usepackage{physics}
\usepackage{mathtools}
\usepackage{soul}
\usepackage{multirow}

\usepackage{enumitem}

\usepackage{algorithm}
\usepackage[noend]{algpseudocode}

\usepackage{mathabx}
\allowdisplaybreaks

\usepackage{hyperref}       %
\hypersetup{
    colorlinks=true,
    linkcolor=blue,
    filecolor=magenta,
    urlcolor=cyan,
}


\usepackage{lineno}

\usepackage{setspace}

\newtheorem{remark}{Remark}
\newtheorem{theorem}{Theorem}
\newtheorem{lemma}{Lemma}

\newtheorem{proposition}{Proposition}

\newif\ifsimflags \simflagstrue

\newcommand{\Iset}{[\![I]\!]}
\newcommand{\Tset}{[\![T]\!]}
\newcommand{\Sset}{[\![S]\!]}

\newcommand{\JiSet}{[\![J_i]\!]}

\providecommand{\R}{\mathbb{R}}

\newcommand{\X}{\mathcal{X}}

\newcommand{\ind}[1]{\mathbf{1}\left\{#1\right\}}

\newcommand{\taij}{\tau_{i,j}^{\rm a}}
\newcommand{\tdij}{\tau_{i,j}^{\rm d}}

\newcommand{\Jgrid}{J^{\text{grid}}}
\newcommand{\Jev}{J^{\text{EV}}}
\newcommand{\Jport}{J^{\text{port}}}
\newcommand{\Jsys}{J^{\text{sys}}}

\newcommand{\probSO}{\hyperref[eq:opt_SO]{\textup{(SO)}}}
\newcommand{\probDone}{\hyperref[eq:D1]{\textup{(D1)}}}
\newcommand{\probSP}{\hyperref[eq:S1-miqp]{\textup{(SP)}}}
\newcommand{\probR}[1]{\hyperref[eq:R#1]{\textup{(R#1)}}}

\newsavebox{\fitcolumnbox}
\newcommand{\fitcolumn}[1]{%
  \sbox{\fitcolumnbox}{#1}%
  \ifdim\wd\fitcolumnbox>\columnwidth
    \resizebox{\columnwidth}{!}{\usebox{\fitcolumnbox}}%
  \else
    \usebox{\fitcolumnbox}%
  \fi}

\title{\LARGE
\textbf{\textsf{MAC+}}: \textbf{\textsf{M}}obility-\textbf{\textsf{A}}ware \textbf{\textsf{C}}o-Expansion \textbf{\textsf{P}}lanning for\\ Distribution Grid and Electric Vehicle Charging Infrastructure
}

\author{Yi Ju$^{1,2,\dagger}$, Lunlong Li$^{3}$, Jingchun Wang$^{1,4}$, and Scott Moura$^{1}$%
\thanks{$^{1}$Department of Civil and Environmental Engineering, University of California, Berkeley.
$^{2}$Harvard John A. Paulson School of Engineering and Applied Sciences.
$^{3}$Department of Civil and Environmental Engineering, the Hong Kong University of Science and Technology.
$^{4}$Department of Civil and Environmental Engineering, University of Illinois Urbana-Champaign.
$^{\dagger}$Corresponding author. {\tt yi\_ju@seas.harvard.edu}.}%
}

\begin{document}

\maketitle
\thispagestyle{plain}
\pagestyle{plain}

\begin{abstract}

We study the least-cost co-expansion of power distribution networks and public charging ports to accommodate the growth of electric vehicles (EVs).
Our approach, \textsf{MAC+}, exploits the flexibility of charging along each EV's itinerary and coordinates it across the fleet to reduce the peak violations of each feeder's hosting capacity and port capacity.

This coordination is an extremely large, spatio-temporally coupled optimization problem with binary plug-in variables.
We solve it by \emph{iterated price response} (IPR), which repeatedly posts a two-tier tariff on feeder load and port occupancy, lets millions of EVs best-respond in parallel, and certifies the optimality gap at every iteration.
We develop a custom exact dynamic-programming algorithm that solves each EV's best-response problem, a mixed-integer quadratic program, about 20 times faster than off-the-shelf solvers.
The case study covers the San Francisco Bay Area at $30\%$ EV adoption, with $1.8$ million EVs on week-long itineraries and $1{,}303$ feeder service regions, each with a utility-published hosting capacity and a port capacity set by its existing public ports.
On this instance, IPR reaches a certified gap below $10\%$ in a few hours.
\textsf{MAC+} saves about \$15 billion of the upgrade investment that unmanaged charging would require.
It also needs over $90\%$ less investment than charging that is fully coordinated but flexible only within each session, and over $80\%$ less than charging that is flexible along the itinerary but uncoordinated.

\end{abstract}

\section{Introduction}\label{sec:intro}

Electrifying personal vehicles moves a large share of transport energy onto the power grid.
Under unmanaged charging, this demand concentrates in the hours and places where vehicles plug in, and accommodating it requires expanding both the public charging infrastructure~\cite{muratori2025trends} and the distribution grid~\cite{li2024impact, elmallah2022can}.
However, EV charging is flexible, since a parked vehicle can often charge later or more slowly without affecting the driver's trips~\cite{garcia2014plug}.
Regions therefore have two options for accommodating this demand: (i) expanding the two infrastructures, and (ii) coordinating charging to increase the utilization of existing capacity.
How much coordination can achieve, and how much expansion remains necessary, is a critical question for regulators, utilities, and infrastructure investors, whose separate decisions add up to one regional outcome.

The answer depends on the interplay of mobility, the distribution network, and the charging infrastructure.
Mobility determines where and when each EV can charge and how much energy its trips require.
The distribution network limits the charging load that each feeder can host, and the charging infrastructure limits how many EVs can plug in at the same time in each area.
Charging that is shifted away from a congested feeder therefore needs both feeder capacity and a free port at another stay of the EV's itinerary, and the shifted energy must still cover the EV's upcoming trips.
Wu et al.~\cite{wu2024planning} find that shifting Bay Area charging away from the evening peak requires more than 18,000 additional fast charging ports.
Chen et al.~\cite{chen2025quantifying} show that grid capacity and charger supply are complementary, so expanding either infrastructure alone produces diminishing returns.
In both studies, however, the charger requirement is evaluated after the optimization rather than imposed as a constraint, so neither model plans the two infrastructures jointly.
We therefore review how large-scale studies of \emph{real-world systems} model these three components and how they compute coordinated charging.
Table~\ref{tab:lit} in Appendix~\ref{appx:lit} summarizes the comparison.

\textbf{Mobility}\quad Most large-scale studies represent mobility only partially, through isolated charging sessions~\cite{xu2018planning, powell2022charging, li2024impact, xu2026proactive}, day-level energy totals~\cite{needell2016potential, xu2023electric, duan2025grid}, or aggregate charging profiles~\cite{wolinetz2018simulating, hanig2025driving, sanvito2026coordinated}.
Among the studies that follow the connected itinerary of each EV, some restrict it to a single day~\cite{chen2025quantifying, zhou2026decentralized}, which rules out shifting charging between days, even though many EVs need to charge only every few days.
Others cover multiple days~\cite{muratori2018impact, wei2021personal, wu2024planning, martin2025cascading, chen2025vehicle}, and some of them let charging shift between the stays of these itineraries~\cite{wu2024planning, martin2025cascading, chen2025vehicle}.
Ju et al.~\cite{ju2026scalable} (hereafter \textsf{MAC}) formalize this \emph{mobility-aware flexibility} as bounds on each EV's cumulative charge along its week-long itinerary, set by the battery capacity and the energy of the upcoming trips.

\textbf{Power infrastructure}\quad Many studies do not model the network and treat a region as one aggregate pool~\cite{li2025unlocking, chen2025quantifying, li2026facilitating, chen2025vehicle}.
For example, Li et al.~\cite{li2025unlocking} coordinate charging to flatten the citywide load of Shenzhen.
However, both transmission~\cite{duan2025grid} and distribution~\cite{muratori2018impact, li2024impact} networks can become congested, which limits the power they can safely deliver.
In particular, the distribution network is where EV charging load, through the charging infrastructure, enters the grid directly.
It therefore congests first when charging concentrates, and its expansion should be planned together with the deployment of charging ports.
Flattening the aggregate load need not relieve the network and can even raise its maximum line flows~\cite{zhou2026decentralized}.
Navidi et al.~\cite{navidi2023coordinating} show that coordinating distributed energy resources to minimize distribution-grid violations reduces the share of transformers with violations in 2050 from 81\% to 28\%.
Xu et al.~\cite{xu2026proactive} evaluate transformer overloads in Bay Area feeders by power-flow simulation, but only after optimizing the charging schedules without the network.
\textsf{MAC} instead coordinates the charging of the whole fleet to minimize the peak violation of every feeder's time-varying hosting capacity.

\textbf{Charging infrastructure}\quad A charging port typically serves one vehicle at a time, so port expansion is driven by the number of simultaneous plug-ins, whereas feeder expansion is driven by the charging power, and the two can conflict.
Spreading charging over many plug-ins at low power relieves the feeders but occupies more ports, and concentrating it in fewer plug-ins does the opposite.
Most studies, including \textsf{MAC}, do not treat charger availability as a limited resource, even implicitly.
They assume that a charger is available wherever an EV parks or, for certain groups of drivers, at home and work~\cite{powell2022charging, li2025unlocking, chen2025vehicle, li2026facilitating, zhou2026decentralized}.
Some studies use a zone's total charger stock, which represents charger capacity only in aggregate~\cite{wu2024planning, sanvito2026coordinated, chen2025quantifying}.
Others use station locations to decide at which stays an EV can charge, but simplify availability to proximity without tracking port occupancy~\cite{xu2018planning, martin2025cascading}.
Tracking port occupancy requires a binary plug-in decision at every stay of every EV, which for a metropolitan fleet means tens of millions of binary variables.

\textbf{Methodology}\quad Studies also differ in how they compute charging schedules, from rule-based simulation and heuristic control to optimization.
At a metropolitan scale, most optimization studies adopt one of two compromises.
The first keeps individual vehicles but drops the coupling between them, optimizing in batches of a few thousand~\cite{li2025unlocking}, community by community~\cite{li2026facilitating}, or vehicle by vehicle against an exogenous signal~\cite{chen2025vehicle, martin2025cascading}.
The second keeps the coupling but merges individual vehicles into aggregate pools~\cite{duan2025grid, sanvito2026coordinated, chen2025quantifying, hanig2025driving}.
Martin et al.~\cite{martin2025cascading} measure the cost of the first compromise.
When millions of EVs respond to the same external signal, their combined response changes the system conditions on which the signal was based, and the intended benefit can disappear.
Under the second compromise, an aggregate pool admits schedules that cannot be split into feasible schedules of the individual EVs.
Chen et al.~\cite{chen2025quantifying} limit this error by a deliberately conservative inner approximation.
Neither compromise can quantify how far its solution lies from the coordinated optimum.
\textsf{MAC} avoids both compromises by coordinating 1.8 million individual EVs through their shared feeder capacities.
Its solution is implemented through posted prices and comes with a certified bound on its distance to the optimum, so planners can weigh the full potential of coordination against capacity expansion.

In this paper, we introduce \textsf{MAC+}, a mobility-aware framework for the co-expansion planning of distribution feeders and public charging ports.
\textsf{MAC+} inherits from \textsf{MAC} the mobility-aware flexibility and the coordination by iterated price response, and it extends them with shared charging ports as a second capacitated resource.
The ports make each plug-in a binary decision, so the per-EV best-response problem becomes mixed-integer and the coordination problem non-convex.
\textsf{MAC+} nevertheless implements its solution through posted prices and certifies its suboptimality.
This paper is self-contained in its technical substance, although \textsf{MAC} gives further detail on the data processing and on the design of the price updates.
The contributions of this paper are:
\begin{itemize}[leftmargin=*]
\item We incorporate mobility-aware flexibility as an operational resource in infrastructure expansion planning. The expansion of distribution feeders and public charging ports is optimized jointly against the capacities of each feeder and of the ports in its service area, rather than evaluated after charging is scheduled.
\item We extend iterated price response to coordinate charging under two capacitated resources. Although the problem is non-convex, we characterize the quality of its solution empirically by an optimality certificate computed at every iteration, and we prove that the relative integrality gap vanishes as the fleet grows. As a key ingredient of the pipeline, a custom exact dynamic-programming solver computes each EV's best response about 20 times faster than off-the-shelf solvers.
\item In a case study of the San Francisco Bay Area with 1.8 million EVs, we compare the expansion needs of both infrastructures across charging policies from unmanaged to fully coordinated, map where these needs arise, trace the co-expansion frontier along the port-to-feeder cost ratio, and assess bidirectional charging and the sensitivity to modeling assumptions.
\end{itemize}

The rest of the paper is organized as follows.
Section~\ref{sec:pro} formulates the coordination problem under the two capacities and states the iterated price response method with its optimality certificate.
Section~\ref{sec:solver} develops and benchmarks the exact solver for per-EV pricing problems.
Section~\ref{sec:res} presents the Bay Area case study, and Section~\ref{sec:conclusion} concludes.

\section{Charging Coordination with Feeder and Charger-Port Capacities}\label{sec:pro}

This section formulates the coordination of an EV fleet under feeder and port capacities and states the iterated price response (IPR) method, which carries an optimality certificate at every iteration.

\subsection{Problem Formulation}\label{sec:pro-form}

We consider a time horizon of $T$ unit-length time slots $t \in \Tset$, one week at hourly resolution in our experiments.
Let $i \in \Iset$ index the EVs.
Each EV follows a fixed itinerary, i.e., a sequence that alternates between trips and stays.
Let $j \in \JiSet$ index the stays of EV $i$.
Stay $j$ occupies the time window $[\taij, \tdij]$, with duration $\tau_{i,j} = \tdij - \taij$, at a location served by feeder $s_{i,j}$.
At each stay, the EV decides whether to plug in and how much energy to charge, collected in the plug-in indicators $\chi_i \in \{0,1\}^{J_i}$ and the charging energies $q_i \in \R^{J_i}_{\ge 0}$, with $\chi_{i,j} = \ind{q_{i,j} > 0}$.
For simplicity, we assume a plugged-in EV charges at a uniform rate over the stay.
The pair $(q_i, \chi_i)$ must lie in a feasible set $\X_i$, constructed in Section~\ref{sec:stay-constr}, which is compact and, because of the binary $\chi_i$, non-convex.

Since stays of different EVs are not aligned in time, we map them onto the common time slots.
The stay-to-time map $A_i \in \R^{T \times J_i}$ records the fraction of each slot that each stay covers,
\begin{equation}\label{eq:Ai}
A_i[t,j] = \abs{[t-1,t] \cap [\taij, \tdij]} \in [0,1],
\end{equation}
and $\overline{A}_i = A_i\, \mathrm{diag}(1/\tau_{i,1}, \dots, 1/\tau_{i,J_i})$ is its column-normalized counterpart, so the slot-level charging profile of EV $i$ is $\overline{A}_i\, q_i$.
In each slot $t$, the fleet draws on two resources of each feeder $s \in \Sset$.
The \emph{charging load} $l_{s,t}$ is the total EV charging power on the feeder, and the \emph{occupancy} $n_{s,t}$ is the number of occupied shared charging ports in its service region.
Both are linear in the per-stay decisions,
\begin{equation}\label{eq:agg-ln}
l = \sum\nolimits_i M^{\rm l}_i\, q_i, \qquad
n = \sum\nolimits_i M^{\rm n}_i\, \chi_i,
\end{equation}
where $M^{\rm l}_i, M^{\rm n}_i \in \R^{ST \times J_i}$ route each stay to its feeder,
\begin{subequations}\label{eq:Mmaps}
\begin{align}
\big(M^{\rm l}_i\big)_{(s,t),\,j} &= \ind{s_{i,j} = s}\, \overline{A}_i[t,j], \\
\big(M^{\rm n}_i\big)_{(s,t),\,j} &= \ind{s_{i,j} = s}\, A_i[t,j],
\end{align}
\end{subequations}
so a plugged-in stay spreads its energy $q_{i,j}$ uniformly over its window and occupies a shared port for the covered fraction of each slot.
For drivers with a private home charger, the columns of $M^{\rm n}_i$ for their home stays are zero.

The charging load on feeder $s$ is limited by its time-varying hosting capacity $C^{\rm l}_{s,t}$, and the occupancy is limited by its port capacity $C^{\rm n}_{s}$.
We penalize the \emph{peak violation} of each capacity, defined as the largest amount by which the aggregate exceeds the capacity over the horizon,
\begin{align}\label{eq:Jsys}
\Jsys(l, n) &= \Jgrid(l) + w_{\rm n}\, \Jport(n) \notag\\
&= \sum\nolimits_s \max_t\, \big[l_{s,t} - C^{\rm l}_{s,t}\big]^+ \notag\\
&\quad + w_{\rm n} \sum\nolimits_s \max_t\, \big[n_{s,t} - C^{\rm n}_{s}\big]^+,
\end{align}
where $[x]^+ = \max\{x, 0\}$.
With the weight of the load violation normalized to $w_{\rm l} = 1$, the weight $w_{\rm n} \ge 0$ is the cost ratio between one shared port and one unit of hosting capacity.
Setting $w_{\rm n} = 0$ recovers a hosting-capacity-only model in which ports are free.

Each EV also carries the disutility\footnote{The second equality holds when each slot intersects at most one stay of the EV. Consecutive stays can share a boundary slot, e.g., departing one location and arriving at another within the same hour, and there the right-hand side, which defines $\Jev_i$, drops the cross term.}
\begin{equation}\label{eq:Jev}
\Jev_i(q_i) = \frac{\kappa}{2}\norm{\overline{A}_i\, q_i}_2^2
= \frac{\kappa}{2} \sum\nolimits_j \alpha_{i,j}\, q_{i,j}^2,
\quad \alpha_{i,j} = \norm{\overline{A}_i[\cdot, j]}_2^2,
\end{equation}
where $\kappa > 0$ is a small constant.
Evaluated on the slot-level charging profile, $\Jev_i$ penalizes high charging power.
The constant $\kappa$ can be read as a battery-wear coefficient, and it also serves as a numerical instrument that makes $\Jev_i$ strongly convex in $q_i$ and thus stabilizes the computation.

The \textbf{social optimum (SO)} of coordinated charging solves
\begin{subequations}\label{eq:opt_SO}
\begin{align}
    ({\rm SO})\quad \min_{q, \chi, l, n}~~ & \Jsys(l, n) + \sum\nolimits_i \Jev_i(q_i)\\
    \text{s.t.}~~ & (q_i, \chi_i) \in \X_i, \quad \forall i \label{eq:SO-Xi}\\
    & \sum\nolimits_{i} M^{\rm l}_i\, q_i - l = 0 \quad (\text{dual:}~\lambda^{\rm l}) \label{eq:SO-share-l}\\
    & \sum\nolimits_{i} M^{\rm n}_i\, \chi_i - n = 0 \quad (\text{dual:}~\lambda^{\rm n}), \label{eq:SO-share-n}
\end{align}
\end{subequations}
and we denote its optimal value by $J^\star$.

\subsection{Construction of the Feasible Set $\X_i$}\label{sec:stay-constr}

The trip after stay $j$ consumes energy $D_{i,j} \ge 0$, obtained from an energy consumption model along the itinerary.
EV $i$ starts with energy $E_{i,0}$ and has battery capacity $E^{\max}_i$.
Four physical requirements define $\X_i$.
\begin{enumerate}[label=\arabic*), leftmargin=*]
\item The \emph{charger power cap} limits the energy of a stay to $\overline{Q}_{i,j} = P^{\rm L2}\, \tau_{i,j}$, with $P^{\rm L2}$ the level-2 charging power.
\item A \emph{meaningful plug-in} must add at least a fraction $\underline{\delta}$ of the battery capacity, $\underline{Q}_{i,j} = \underline{\delta}\, E^{\max}_i$.
\item The \emph{battery capacity} caps the stored energy at $E^{\max}_i$.
\item \emph{Mobility adequacy} requires the stored energy to stay nonnegative during all trips and to meet any prescribed departure targets.
\end{enumerate}
\begin{remark}[Plug-in floor]\label{rmk:floor}
The floor of requirement 2) plays two roles.
First, it refines the behavioral model.
Since finding a charger and plugging in is an effort, drivers plug in only for a substantial charge.
We model this effort by a floor rather than by a fixed cost per plug-in, as we expect a floor to be more easily calibrated from empirical surveys and measurements.
Second, the floor is where the binary plug-in decision enters the feasible set.
Without it, the charge of a stay may take any value in $[0, \overline{Q}_{i,j}]$.
With it, the charge is either zero or at least $\underline{Q}_{i,j}$, and $\chi_{i,j}$ is the well-defined indicator of which option is taken.
\end{remark}
The energy at the end of stay $j$ is $e_{i,j} = E_{i,0} + \sum_{j' \le j} q_{i,j'} - \sum_{j' < j} D_{i,j'}$, and requirements 3) and 4) read $e_{i,j} \le E^{\max}_i$ and
$e_{i,j} \ge \max\big\{\underline{E}^{\rm dep}_{i,j},\ \underline{E}^{\rm arr}_{i,j+1} + D_{i,j}\big\}$,
where $\underline{E}^{\rm dep}_{i,j}$ is an optional energy target at departure from stay $j$ and $\underline{E}^{\rm arr}_{i,j}$ is a safety threshold on the energy at arrival at stay $j$, both $0$ by default, with $\underline{E}^{\rm arr}_{i,J_i+1} = 0$.
Let $H_i \in \R^{J_i \times J_i}$ be the cumulative-sum matrix, $(H_i q_i)_j = \sum_{j' \le j} q_{i,j'}$.
Substituting $e_{i,j}$ yields sandwich bounds on the cumulative charge,
\begin{align}
& \underline{S}_{i,j} = \max\big\{\underline{E}^{\rm dep}_{i,j},\ \underline{E}^{\rm arr}_{i,j+1} + D_{i,j}\big\} - E_{i,0} + \sum\nolimits_{j' < j} D_{i,j'}, \notag\\
& \overline{S}_{i,j} = E^{\max}_i - E_{i,0} + \sum\nolimits_{j' < j} D_{i,j'}, \label{eq:S-bounds}
\end{align}
and the feasible set is
\begin{align}\label{eq:Xi}
\X_i = \big\{ (q_i, \chi_i) \in \R^{J_i} \times \{0,1\}^{J_i}:~ & \chi_i \odot \underline{Q}_i \preceq q_i \preceq \chi_i \odot \overline{Q}_i, \notag\\
& \underline{S}_i \preceq H_i\, q_i \preceq \overline{S}_i \big\},
\end{align}
where $\odot$ denotes element-wise multiplication.
For every stay $j$ with $\underline{Q}_{i,j} > 0$, the indicator constraint in \eqref{eq:Xi} is equivalent to the semi-continuous disjunction
\begin{equation}\label{eq:disjunction}
q_{i,j} \in \{0\} \cup [\underline{Q}_{i,j}, \overline{Q}_{i,j}], \qquad
\chi_{i,j} = \ind{q_{i,j} > 0},
\end{equation}
which makes $\chi_{i,j}$ well defined.
We call the excluded open interval $(0, \underline{Q}_{i,j})$ the \emph{hole} of stay $j$.
Relaxing the disjunction to $q_{i,j} \in [0, \overline{Q}_{i,j}]$ gives the \emph{continuous relaxation} $\X^{\rm cont}_i \supseteq \X_i$.
Not every itinerary admits a nonempty $\X_i$ under the plug-in floor, and we assume $\X_i \neq \emptyset$ in this section.

We restrict attention to uni-directional charging, $q_i \succeq 0$, and Appendix~\ref{appx:v2g} extends the model to bidirectional charging (V2G), whose disjunction gains a mirrored discharge interval.

\subsection{Iterated Price Response and Optimality Certificate}\label{sec:ipr}

\begin{figure}[htbp]
    \centering
    \includegraphics{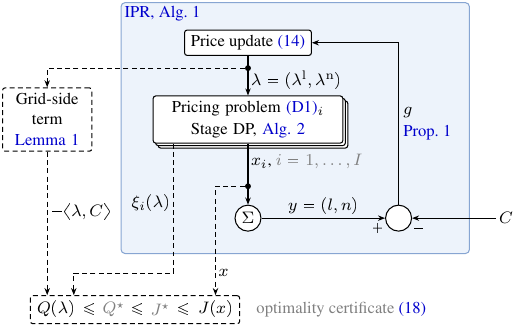}
    \caption{\textbf{Iterated price response (IPR) and its optimality certificate}
    \scriptsize\quad
    Solid arrows show one IPR iteration, and dashed arrows show the certificate.
    The stacked box represents the pricing problems of the $I$ EVs, which are solved by an exact DP detailed in Section~\ref{sec:s1-dp}.
    }
    \label{fig:ipr-diagram}
\end{figure}

Four complexities make \probSO{} hard to solve.
\begin{enumerate}[label=\arabic*), leftmargin=*]
\item \emph{Scale}: the fleet comprises up to millions of EVs, each with its own itinerary.
\item \emph{Spatial coupling}: EVs share the capacities of the feeders they visit, and each EV visits several feeders.
\item \emph{Temporal coupling}: the battery of each EV links its stays over the horizon.
\item \emph{Binary commitment}: the plug-in decisions make every $\X_i$ non-convex.
\end{enumerate}
We address the first two complexities here by pricing the shared capacities.
Under posted prices on the feeder load and the port occupancy, each EV pays for the capacity it uses rather than competing for it with the other EVs.
Problem \probSO{} then decomposes into independent per-EV pricing problems, which are solved in parallel.
The spatial coupling enters only through the prices, which are updated from one iteration to the next (Fig.~\ref{fig:ipr-diagram}).
The last two complexities remain inside each pricing problem, which Section~\ref{sec:solver} solves exactly.

Write $x_i = (q_i, \chi_i)$, $M_i = \mathrm{diag}(M^{\rm l}_i, M^{\rm n}_i)$, $y = (l, n)$, and $C = (C^{\rm l}, C^{\rm n})$ with $C^{\rm n}_{s,t} \equiv C^{\rm n}_s$, so that the coupling constraints of \probSO{} read $\sum_i M_i x_i = y$.
Let $J(x)$ denote the objective of \probSO{} at a per-EV feasible $x = (x_i)_i$ and their aggregation $y = \sum_i M_i x_i$.
Dualizing the coupling constraints with multipliers $\lambda = (\lambda^{\rm l}, \lambda^{\rm n})$, $\lambda^{\rm l}, \lambda^{\rm n} \in \R^{S \times T}$, gives the dual function
\begin{align}\label{eq:Lag}
Q(\lambda) &\coloneqq \min_{\mathclap{x \in \prod_i \X_i,\, y}}\ \Jsys(y) +\sum\nolimits_i \Jev_i(q_i) + \Big\langle \lambda, \sum\nolimits_i M_i x_i - y \Big\rangle \notag\\
&= \sum\nolimits_i \xi_i(\lambda) + \min\nolimits_y \big\{ \Jsys(y) - \langle \lambda, y \rangle \big\},
\end{align}
which separates over the EVs into
\begin{equation}\label{eq:D1}
({\rm D1})_i\quad \xi_i(\lambda) \coloneqq \min_{x_i \in \X_i}\ \Jev_i(q_i) + \langle \lambda, M_i x_i \rangle .
\end{equation}
We call \probDone{}$_i$ the \emph{pricing problem} of EV $i$.
Its solution is the exact best response of EV $i$ to a posted \emph{two-tier tariff}, which consists of a load price $\lambda^{\rm l}_{s,t}$ per unit of energy and a port price $\lambda^{\rm n}_{s,t}$ per occupied port-slot.
Under such a tariff, an EV plugs in at a stay only if the benefit of charging there exceeds the port charge for the whole stay.
Two-part tariffs of this kind are proposed ad hoc in practice, and here they arise as the multipliers of the coupling constraints.

The grid-side term $\min\nolimits_y \left\{ \Jsys(y) - \langle \lambda, y \rangle \right\}$ admits a closed form, separately for each resource.
\begin{lemma}[Capped simplices]\label{lem:Lambda}
For a resource with weight $w \ge 0$ and capacity profiles $C_s \in \R^T$, let $\Lambda(w) \coloneqq \{\lambda \in \R^{S \times T} : \lambda_s \succeq 0,\ \sum_t \lambda_{s,t} \le w\ \ \forall s\}$. Then
\begin{equation*}
\min_{u}\ w \sum\nolimits_s \max\nolimits_t\, [u_{s,t} - C_{s,t}]^+ - \langle \lambda, u \rangle
= -\langle \lambda, C \rangle - \mathbb{I}_{\Lambda(w)}(\lambda),
\end{equation*}
where $\mathbb{I}_{\Lambda(w)}$ is $0$ on $\Lambda(w)$ and $+\infty$ elsewhere.
\end{lemma}
\begin{proof}
See Appendix~\ref{appx:gap}.
\end{proof}

On the \emph{dual-feasible set} $\Lambda \coloneqq \Lambda(1) \times \Lambda(w_{\rm n})$, the dual function reads
\begin{equation}\label{eq:Q-Lambda}
Q(\lambda) = \sum\nolimits_i \xi_i(\lambda) - \langle \lambda, C \rangle ,
\end{equation}
and outside $\Lambda$ it equals $-\infty$.

By weak duality, $Q(\lambda) \le J^\star$ for every $\lambda$.
The tightest of these lower bounds is the dual optimum $Q^\star \coloneqq \max_{\lambda \in \Lambda} Q(\lambda)$.
A natural way to approach $Q^\star$ is projected supergradient ascent,
\begin{equation}\label{eq:pga}
\lambda \gets \Pi_\Lambda\big[\lambda + \eta\, g\big],
\end{equation}
where $\eta > 0$ is a step size, $\Pi_\Lambda$ is the Euclidean projection onto $\Lambda$, and $g$ is a supergradient of $Q$ at $\lambda$, i.e., $Q(\lambda') \le Q(\lambda) + \langle g, \lambda' - \lambda \rangle$ for all $\lambda' \in \Lambda$.
For $\lambda \in \Lambda$, let $x_i(\lambda)$ be an exact solution of \probDone{}$_i$, chosen arbitrarily if several plug-in patterns tie, and let $y(\lambda) \coloneqq \sum_i M_i x_i(\lambda)$ be the \emph{deployed} aggregates.
\begin{proposition}[Supergradient]\label{prop:supergrad}
For every $\lambda \in \Lambda$,
$g(\lambda) \coloneqq y(\lambda) - C$
is a supergradient of $Q$ at $\lambda$.
\end{proposition}
\begin{proof}
See Appendix~\ref{appx:gap}.
\end{proof}

We call $g(\lambda)$ the \emph{deployed shortage}, because it is the amount by which the aggregates of the best responses exceed the capacities.
Step \eqref{eq:pga} with $g = g(\lambda)$ thus raises the prices where the deployed aggregates exceed the capacities and lowers them elsewhere.

This ascent is the \emph{iterated price response} (IPR) method, which the companion paper \textsf{MAC}~\cite{ju2026scalable} originally developed for convex charging coordination.
IPR also applies to \probSO{} as Proposition~\ref{prop:supergrad} requires no convexity.
Each IPR iteration repeats two steps.
In the first step, the tariff $\lambda$ is posted, and every EV solves its pricing problem \probDone{}$_i$, all in parallel.
In the second step, the tariff is updated from the deployed shortage $g(\lambda)$.
IPR refines the plain step \eqref{eq:pga} with three ingredients.
First, it replaces the Euclidean projection by an entropic mirror step~\cite[Ch.~9]{beck2017first} for each feeder $s$ and resource $r \in \{{\rm l}, {\rm n}\}$,
\begin{equation}\label{eq:mw}
\lambda^{r}_s \gets \Pi_{\Lambda(w_r)}\Big[ \lambda^{r}_s \odot \exp\big( \eta\, g^{r}_s / c^{r}_s \big) \Big], \quad
c^{r}_s = \max\{ \|g^{r}_s\|_\infty, c_0 \},
\end{equation}
where $g^r_s$ is the block of $g(\lambda)$ for feeder $s$ and resource $r$, $c_0 > 0$ is a small floor, and $\Pi_{\Lambda(w_r)}$ rescales the price mass to at most $w_r$.
Second, it halves $\eta$ until the step does not lower $Q$.
Third, it starts from the softmax price
\begin{equation}\label{eq:lse}
\widehat{\lambda}^{r}_{s,t} = w_r\, \frac{\exp\big(\beta\, \delta^r_{s,t} / \overline{\delta}^r_s\big)}{\sum_{\tau} \exp\big(\beta\, \delta^r_{s,\tau} / \overline{\delta}^r_s\big)}
\end{equation}
on every feeder with $\overline{\delta}^r_s \coloneqq \max_t \delta^r_{s,t} > 0$ and from a zero price on all other feeders, where $\delta^r_s = \hat y^r_s - C^r_s$ is the shortage profile of the unpriced aggregates $\hat y = \sum_i M_i \arg\min_{x_i \in \X_i} \Jev_i(q_i)$ and $\beta > 0$ is a sharpness parameter.
We refer to~\cite{ju2026scalable} for the motivation of these ingredients and for safeguards against the absorbing zero of multiplicative updates.
Algorithm~\ref{alg:ipr} summarizes IPR.

\begin{algorithm}
\caption{Iterated price response (IPR)}\label{alg:ipr}
\begin{algorithmic}[1]\small
\Procedure{IPR}{$C, \beta, \eta_0, K$}
\State $\hat y \gets \sum_i M_i \arg\min_{x_i \in \X_i} \Jev_i(q_i)$
\Comment{unpriced sweep}
\State $\lambda^{r} \gets \widehat{\lambda}^{r}$ by \eqref{eq:lse}, $r \in \{{\rm l}, {\rm n}\}$
\Comment{shortage seed}
\State $\eta \gets \eta_0$
\For{$k = 0, 1, \dots, K-1$}
\State $x_i(\lambda) \gets$ solve \probDone{}$_i$, $\forall i$
\Comment{best responses}
\State $g \gets y(\lambda) - C$
\Comment{supergradient, Prop.~\ref{prop:supergrad}}
\State keep $x^{\rm best}$ of least $J(x(\lambda))$ and $\lambda^{\rm best}$ of largest $Q(\lambda)$
\State $\lambda^+ \gets$ update \eqref{eq:mw} for every $s$ and $r$
\State backtrack $\eta \gets \eta/2$ until $Q(\lambda^+) \ge Q(\lambda)$; $\lambda \gets \lambda^+$
\EndFor
\State \Return $(x^{\rm best}, \lambda^{\rm best})$ and the gap $J(x^{\rm best}) - Q(\lambda^{\rm best})$
\EndProcedure
\end{algorithmic}
\end{algorithm}

Every dual function is concave, so step \eqref{eq:pga} with suitably diminishing step sizes drives the best dual value to $Q^\star$~\cite[Ch.~8]{beck2017first}.
IPR trades this guarantee for nondecreasing dual values, since its backtracking can stall at a kink of $Q$ where the chosen supergradient is not an ascent direction.

Because the sets $\X_i$ are non-convex, $Q^\star$ can lie below $J^\star$.
Theorem~\ref{thm:gap} shows that this integrality gap is small relative to $J^\star$ at scale.

\begin{theorem}[Vanishing relative integrality gap]\label{thm:gap}
Let $d \coloneqq 2ST$ be the number of coupling constraints in \probSO{}.
Then
\begin{equation}\label{eq:gap-bound}
J^\star - Q^\star \;\le\; (d + 1)\, \max_i\, (\omega_i + b_i) ,
\end{equation}
where $\omega_i \coloneqq \max_{\X_i} \Jev_i - \min_{\X_i} \Jev_i$ is the range of the disutility of EV $i$ over $\X_i$, and
\begin{equation}\label{eq:footprint}
b_i \coloneqq \max_{x, x' \in \X_i}\ \sum\nolimits_s \Big( \big\| M^{\rm l}_i (q - q') \big\|_{s,\infty} + w_{\rm n} \big\| M^{\rm n}_i (\chi - \chi') \big\|_{s,\infty} \Big)
\end{equation}
is its weighted footprint.
Here $\|\cdot\|_{s,\infty}$ is the maximum over the slots of feeder $s$.
In particular, if $\omega_i$ and $b_i$ are bounded uniformly in $I$ while $J^\star \ge c\, I$ for some $c > 0$, the relative gap $(J^\star - Q^\star)/J^\star$ vanishes as $I \to \infty$ at fixed $S$ and $T$.
\end{theorem}
\begin{proof}
Appendix~\ref{appx:gap} applies the Shapley--Folkman lemma~\cite{starr1969quasi} as in the unit-commitment argument of~\cite{bertsekas1983optimal}.
\end{proof}

The footprint $b_i$ bounds the change in the penalty that one EV's decision can cause.
The range $\omega_i$ depends only on $\kappa$ and the EV's stays, and the footprint $b_i$ depends only on the level-2 power, the number of its stays, and the few feeders it visits.
Both therefore stay bounded as the fleet grows, while $J^\star$ grows in proportion to $I$ once the fleet outgrows the fixed capacities, which is the regime where expansion is needed.

Theorem~\ref{thm:gap} bounds the integrality gap a priori, and IPR complements it with an a posteriori optimality certificate at every iteration.
Each best response lies in $\X_i$, so every deployed response $x(\lambda)$ is feasible for \probSO{}, and $J(x(\lambda))$ is an upper bound on $J^\star$.
By weak duality, which does not require convexity, $Q(\lambda)$ is a lower bound on $J^\star$.
Together,
\begin{equation}\label{eq:weak}
Q(\lambda) \;\le\; J^\star \;\le\; J(x), \qquad \forall\, x \in \textstyle\prod_i \X_i,\ \ \forall\, \lambda .
\end{equation}
The gap $J(x^{\rm best}) - Q(\lambda^{\rm best})$ of the best certified pair is therefore known at every iteration.
It bounds both the suboptimality of $x^{\rm best}$ and the integrality gap $J^\star - Q^\star$, and $x^{\rm best}$ is deployable, since it is the fleet's response to a posted tariff.
IPR can stop once its best certified pair $(x, \lambda)$ meets a relative tolerance $\epsilon^{\rm rel} > 0$,
\begin{equation}\label{eq:rel-gap}
J(x) - Q(\lambda) \;\le\; \epsilon^{\rm rel} \max\{1, |J(x)|, |Q(\lambda)|\}.
\end{equation}
Note that the lower bound requires the exact values $\xi_i(\lambda)$, since an inexact solve overstates them.

\section{Per-EV Subproblem Solvers}\label{sec:solver}

Each iteration of Algorithm~\ref{alg:ipr} solves the pricing problem \probDone{} for every EV (the stacked box in Fig.~\ref{fig:ipr-diagram}), potentially millions of mixed-integer instances per sweep, too many for a generic solver.
This section develops faster solvers that exploit the structure of \probDone{}.

Throughout this section we fix one EV and drop the index $i$.
By \eqref{eq:Mmaps}, the price term of \probDone{} is $\langle \lambda, M x \rangle = \sum_j (\ell_j q_j + \pi_j \chi_j)$, with the stay prices
\begin{equation}\label{eq:stay-prices}
\ell_j = \sum\nolimits_t \lambda^{\rm l}_{s_j,t}\, \overline{A}[t,j], \qquad
\pi_j = \sum\nolimits_t \lambda^{\rm n}_{s_j,t}\, A[t,j],
\end{equation}
the load price averaged over stay $j$ and the port price summed over its slots (each slot weighted by its covered fraction).
A home stay with a private charger occupies no shared port and has $\pi_j = 0$. In addition, $\pi \equiv 0$ when $w_{\rm n} = 0$.
With \eqref{eq:Jev} and \eqref{eq:Xi}, \probDone{} (hereafter \probSP{}) reads
\begin{align}\label{eq:S1-miqp}
({\rm SP})\quad \min_{q,\, \chi}~~ & \Phi(q, \chi) \coloneqq \sum\nolimits_j \Big[ \frac{c_j}{2}\, q_j^2 + \ell_j\, q_j + \pi_j\, \chi_j \Big] \\
\text{s.t.}~~ & \underline{Q}_j \le q_j \le \overline{Q}_j \ \text{if}\ \chi_j = 1, \quad q_j = 0 \ \text{if}\ \chi_j = 0, \notag\\
& \underline{S}_j \le \sum\nolimits_{j' \le j} q_{j'} \le \overline{S}_j, \quad \chi_j \in \{0, 1\}, \quad \forall j,\notag
\end{align}
with curvature $c_j \coloneqq \kappa\, \alpha_j$ and floor $\underline{Q}_j > 0$.
Section~\ref{sec:feas-oracle} characterizes the feasible set $\X$ of \probSP{}, Section~\ref{sec:s1-dp} solves \probSP{} exactly by a dynamic program, Section~\ref{sec:restrict} compares it with two restricted formulations that admit simpler algorithms, and Section~\ref{sec:solver-bench} benchmarks the solvers.

\subsection{Exact Feasibility Test}\label{sec:feas-oracle}

For a set $A \subseteq \R$ and an interval $[a, b]$, let $A \oplus [a, b] \coloneqq \{x + y : x \in A,\ y \in [a, b]\}$ and $A \ominus [a, b] \coloneqq A \oplus [-b, -a]$, and let $[\underline{S}_0, \overline{S}_0] = \R$.
Define the \emph{forward reachable sets} $\mathcal{R}_j$ by
\begin{align}\label{eq:Rj}
& \mathcal{R}_0 = \{0\}; \qquad \text{for}~~ j = 1, \dots, J:\\
& \mathcal{R}_j = [\underline{S}_j, \overline{S}_j] \cap \big( \mathcal{R}_{j-1} \cup (\mathcal{R}_{j-1} \oplus [\underline{Q}_j, \overline{Q}_j]) \big). \notag
\end{align}
Define the \emph{backward co-reachable sets} $\mathcal{B}_j$ by
\begin{align}\label{eq:Bj}
& \mathcal{B}_J = [\underline{S}_J, \overline{S}_J]; \qquad \text{for}~~ j = J-1, \dots, 0:\\
& \mathcal{B}_j = [\underline{S}_j, \overline{S}_j] \cap \big(\mathcal{B}_{j+1}
 \cup (\mathcal{B}_{j+1} \ominus [\underline{Q}_{j+1}, \overline{Q}_{j+1}]) \big).\notag
\end{align}
The unshifted and shifted sets are the no-charge and charge branches of stay $j$ in \eqref{eq:Rj} and of stay $j+1$ in \eqref{eq:Bj}.
Thus $\mathcal{R}_j$ collects the cumulative charges after stay $j$ that the stays \emph{up to} $j$ can reach within their bounds, and $\mathcal{B}_j$ collects those from which the stays \emph{after} $j$ can be completed within theirs.

A shift by an interval, a union, and an intersection with an interval each map a finite union of closed intervals to a finite union of closed intervals.
Every $\mathcal{R}_j$ and $\mathcal{B}_j$ is therefore a finite union of disjoint closed intervals, which can be stored exactly as a sorted list of endpoints.

\begin{theorem}[Exact feasibility test]\label{thm:oracle}
\begin{enumerate}[label=(\roman*), leftmargin=*]
\item For $j = 0, \dots, J$, $\mathcal{R}_j$ is the set of cumulative charges $\sum_{k \le j} q_k$ of all $(q_k, \chi_k)_{k \le j}$ that satisfy the constraints of \probSP{} on the stays up to $j$. The set $\mathcal{B}_j$ consists of the $c \in [\underline{S}_j, \overline{S}_j]$ for which some $(q_k, \chi_k)_{k > j}$ satisfies the constraints of \probSP{} on the stays after $j$, with $c + \sum_{k'=j+1}^{k} q_{k'}$ as the cumulative charge after stay $k$.
\item $\X \neq \emptyset \iff \mathcal{R}_J \neq \emptyset \iff 0 \in \mathcal{B}_0$.
\item With $m$ the maximum number of intervals in any $\mathcal{R}_j$ or $\mathcal{B}_j$, either test costs $\mathcal{O}(J m)$.
\end{enumerate}
\end{theorem}
\begin{proof}
See Appendix~\ref{appx:pf-oracle}.
\end{proof}

Intuitively, $m$ stays small: a reachable interval $[a, b]$ and its charged copy $[a + \underline{Q}_j, b + \overline{Q}_j]$ overlap unless $\underline{Q}_j > b - a$, so holes open only next to intervals narrower than the floor, while each charge widens an interval by $\overline{Q}_j - \underline{Q}_j$.

\begin{remark}[Convex tests do not decide feasibility]\label{rmk:cont-insufficient}
Since $\X$ is contained in its continuous relaxation $\X^{\rm cont}$ of Section~\ref{sec:stay-constr}, feasibility of $\X^{\rm cont}$ is necessary for $\X \neq \emptyset$. However, it is not sufficient. A minimal counterexample:
For $J = 1$, the data $[\underline{Q}_1, \overline{Q}_1] = [20, 30]$ and $\underline{S}_1 = \overline{S}_1 = 10$ force $q_1 = 10$, which lies in $[0, 30]$ but in the hole $(0, 20)$, so $\X^{\rm cont} \neq \emptyset$ while $\X = \emptyset$.
\end{remark}

\begin{remark}[Feasibility repair]\label{rmk:curtail-model}
An EV with $0 \notin \mathcal{B}_0$ cannot serve its driving demand under the plug-in floor. For such instances ($< 1\%$ in the case study fleet), we curtail their demand minimally to restore feasibility in data preprocessing.
Let $\delta_j$ be the energy curtailed from the trip after stay $j$.
The repair imposes the requirements of Section~\ref{sec:stay-constr} on the stored energy $e_j$ at the end of stay $j$, with $D_j$ replaced by $D_j - \delta_j$,
\begin{align}\label{eq:curtail}
\min_{q,\, \delta}~~ & \sum\nolimits_j \delta_j\\
\text{s.t.}~~ & q_j \in \{0\} \cup [\underline{Q}_j, \overline{Q}_j], \quad \delta_j \in [0, D_j], \quad \forall j, \notag\\
& e_j = E_0 + \sum\nolimits_{j' \le j} q_{j'} - \sum\nolimits_{j' < j} (D_{j'} - \delta_{j'}), \quad \forall j, \notag\\
& \max\big\{\underline{E}^{\rm dep}_j,\ \underline{E}^{\rm arr}_{j+1} + D_j - \delta_j\big\} \le e_j \le E^{\max}, \quad \forall j,\notag
\end{align}
and the EV keeps the curtailed demands $D_j - \delta^\star_j$ thereafter.
The curtailed energy is unserved mobility.
Solving \eqref{eq:curtail} is an application of the stage DP of Section~\ref{sec:s1-dp}.
\end{remark}

\subsection{Exact Solver: Piecewise-Quadratic Stage DP}\label{sec:s1-dp}

Let $C_j \coloneqq \sum_{j' \le j} q_{j'}$ be the cumulative charge after stay $j$, with $C_0 = 0$.
Stays interact in \probSP{} only through $C_j$, so \probSP{} admits a dynamic program (DP) with a one-dimensional continuous state.
For $j = 1, \dots, J$, let $V_j(C)$ be the optimal value of \probSP{} over the stays from $j$ on when the stays before $j$ have charged $C$,
\begin{align}\label{eq:Vj}
& V_j(C) \coloneqq \min_{(q_k, \chi_k)_{k \ge j}}~ \sum\nolimits_{k \ge j} \Big[ \frac{c_k}{2}\, q_k^2 + \ell_k\, q_k + \pi_k\, \chi_k \Big]\\
& \text{s.t.}~~ \underline{Q}_k \le q_k \le \overline{Q}_k \ \text{if}\ \chi_k = 1, \quad q_k = 0 \ \text{if}\ \chi_k = 0, \notag\\
& \phantom{\text{s.t.}~~} \underline{S}_k \le C + \sum\nolimits_{k'=j}^{k} q_{k'} \le \overline{S}_k, \quad \chi_k \in \{0, 1\}, \quad \forall k \ge j,\notag
\end{align}
with $\min \emptyset = +\infty$ and $V_{J+1} \equiv 0$, so $V_1(0)$ is the optimal value of \probSP{}.
By the principle of optimality, separating the decision at stay $j$ from the decisions at the later stays yields the backward recursion, for $j = J, \dots, 1$,
\begin{align}
V_j(C) &=  \min\big\{ V^{\rm off}_j(C),\ V^{\rm on}_j(C) \big\},~~\text{where} \label{eq:dp}\\
&V^{\rm off}_j(C) =
\begin{cases}
V_{j+1}(C), & C \in [\underline{S}_j, \overline{S}_j],\\
+\infty, & \text{otherwise},
\end{cases} \notag\\
&V^{\rm on}_j(C) = \min_{q_j}
\Big[\frac{c_j}{2} q_j^2 + \ell_j q_j + \pi_j + V_{j+1}(C + q_j)\Big] \notag\\
& \qquad\qquad\quad \text{s.t.}~~q_j \in [\underline{Q}_j, \overline{Q}_j], ~~ C + q_j \in [\underline{S}_j, \overline{S}_j]. \notag
\end{align}
The two terms are the costs of the no-charge and the charge branch of stay $j$, each followed by an optimal completion of the stays after $j$.

Proposition~\ref{prop:dom} shows that the DP is the backward recursion \eqref{eq:Bj} with a cost-to-go attached to each set.
\begin{proposition}\label{prop:dom}
For $j = 1, \dots, J+1$, $\operatorname{dom} V_j \cap [\underline{S}_{j-1}, \overline{S}_{j-1}] = \mathcal{B}_{j-1}$.
In particular, $V_1(0) < \infty$ if and only if $\X \neq \emptyset$.
\end{proposition}
\begin{proof}
See Appendix~\ref{appx:pf-dom}.
\end{proof}

\noindent\textbf{Computability}\quad
Even with a one-dimensional state, a DP over a continuous state is usually solved approximately on a discretized grid, because its cost-to-go functions need not admit a finite representation.
However, the recursion \eqref{eq:dp} needs no discretization, because it is closed on a finitely representable function class.

A function $V : \R \to \R \cup \{+\infty\}$ is \emph{piecewise quadratic} (PWQ) if it is described by a finite set $\Theta$ of \emph{arcs}.
Each arc is a 5-tuple $\theta = (\underline{\theta}, \overline{\theta}, \theta_2, \theta_1, \theta_0)$ that carries the convex quadratic $\phi_\theta(C) = \tfrac{\theta_2}{2} C^2 + \theta_1 C + \theta_0$, with curvature $\theta_2 \ge 0$, on the interval $[\underline{\theta}, \overline{\theta}]$, and
\begin{equation}\label{eq:Theta}
V(C) =
\begin{cases}
\phi_\theta(C), & \text{if } C \in [\underline{\theta}, \overline{\theta}] \text{ for an arc } \theta \in \Theta,\\
+\infty, & \text{if } C \text{ lies in no arc interval}.
\end{cases}
\end{equation}
The arc intervals are sorted and overlap only at their endpoints, where $V$ takes the smaller of the two values.
The base case, $V_{J+1} \equiv 0$, is described by the single arc $(-\infty, +\infty, 0, 0, 0)$.

\begin{proposition}[PWQ invariance]\label{prop:pwq}
Every $V_j$ is PWQ.
Its arc set $\Theta_j$ is computed from $\Theta_{j+1}$ in closed form by the \emph{backward map} $\mathcal{T}_j$ of stay $j$, $\Theta_j = \mathcal{T}_j(\Theta_{j+1})$, which evaluates \eqref{eq:dp} on arc sets.
\end{proposition}
\begin{proof}
See Appendix~\ref{appx:pwq}, which constructs $\mathcal{T}_j$ as Algorithm~\ref{alg:dpstep}.
\end{proof}

Every arc is convex although $V_j$ is not, and the non-convexity of the disjunction enters only through the two-branch minimum.
When $\underline{\delta} = 0$ and $\pi \equiv 0$, the charge branch includes $q_j = 0$ at no extra cost, so $V_j = V^{\rm on}_j$, every $V_j$ is convex, and \probSP{} reduces to the convex per-EV problem of \textsf{MAC}~\cite{ju2026scalable} at the stay level.

Overall, Algorithm~\ref{alg:dp}, the exact DP solver of \probSP{}, runs the backward pass to construct the arc sets $\Theta_J, \dots, \Theta_1$ and then the forward pass to recover an optimal solution $(q, \chi)$.

\begin{algorithm}
\caption{Stage DP for \probSP{}}\label{alg:dp}
\begin{algorithmic}[1]\small
\Procedure{StageDP}{$\{\mathcal{B}_j\}_{j=0}^J$}
\State $\Theta_{J+1} \gets \{(-\infty, +\infty, 0, 0, 0)\}$
\Comment{$V_{J+1} \equiv 0$}
\For{$j = J, \dots, 1$}
\Comment{backward pass}
\State $\Theta_j \gets \mathcal{T}_j(\Theta_{j+1})$
\Comment{Alg.~\ref{alg:dpstep}}
\EndFor
\State $C \gets 0$
\For{$j = 1, \dots, J$}
\Comment{forward pass}
\State $q^{\rm on} \gets$ a minimizing $q_j$ of $V^{\rm on}_j(C)$
\If{$V^{\rm off}_j(C) \le V^{\rm on}_j(C)$}
\State $(q_j, \chi_j) \gets (0, 0)$
\Else
\State $(q_j, \chi_j) \gets (q^{\rm on}, 1)$
\EndIf
\State $C \gets C + q_j$
\EndFor
\State \Return $(q, \chi)$ and $V_1(0)$
\EndProcedure
\end{algorithmic}
\end{algorithm}

\begin{theorem}[Exactness]\label{thm:dp-exact}
The backward pass of Algorithm~\ref{alg:dp} computes the optimal value $V_1(0)$ of \probSP{}.
If $\X \neq \emptyset$, its forward pass recovers an optimal solution $(q, \chi)$ of \probSP{}.
\end{theorem}
\begin{proof}
See Appendix~\ref{appx:pf-exact}.
\end{proof}

With $P \coloneqq \max_j |\Theta_j|$ the maximum number of arcs, each backward map costs $\mathcal{O}(P \log P)$, so the backward pass costs $\mathcal{O}(J P \log P)$ and the forward pass $\mathcal{O}(J P)$.
No polynomial bound on $P$ is known, but $P$ stays small in practice.
On the benchmark EVs of Section~\ref{sec:solver-bench}, our implementation needs at most $19$ arcs per stage for the median EV and at most $64$ for more than $98\%$ of the EVs.
For the remaining EVs, a piece cap merges adjacent arcs to keep $P \le 64$, and Appendix~\ref{appx:cap} shows that the certificate of Section~\ref{sec:ipr} remains valid under this cap.

\subsection{Restricted Pricing Problems}\label{sec:restrict}

We compare three formulations of the pricing problem.
The first is \probSP{} itself, denoted (R0) here:
\begin{equation}\label{eq:R0}
({\rm R0})\quad \min_{q,\, \chi}\ \big\{ \Phi(q, \chi) : (q, \chi) \in \X \big\},
\end{equation}
solved exactly by the stage DP of Section~\ref{sec:s1-dp}.
The second fixes the plug-in pattern at $\chi^\circ$,
\begin{equation}\label{eq:R1}
({\rm R1})\quad \min_{q}\ \big\{ \Phi(q, \chi^\circ) : (q, \chi^\circ) \in \X \big\},
\end{equation}
a convex quadratic program solved by the projected dual ascent of~\cite{ju2026scalable}.
We may use \probR{1} when we expect the plug-in patterns not to change under slight price changes.
The third fixes the stay energies at $q^\circ$ (thus $\chi^\circ$) and optimizes the energy $p_{j,t} \ge 0$ charged in slot $t$ during stay $j$,
\begin{align}\label{eq:R2}
({\rm R2})\quad \min_{p \ge 0}~~ & \sum\nolimits_{j,t} \Big[ \frac{\kappa}{2}\, p_{j,t}^2 + \lambda^{\rm l}_{s_j,t}\, p_{j,t} \Big] \notag\\
\text{s.t.}~~ & \sum\nolimits_t p_{j,t} = q^\circ_j, \quad p_{j,t} \le \overline{A}[t,j]\, \overline{Q}_j, \quad \forall j,
\end{align}
which separates into one quadratic knapsack per stay, solved by bisection on the multiplier of its energy constraint.
We may use \probR{2} to refine a solution of \probR{0} on long stays over which the load price varies in time.

Only \probR{0} returns the exact value $\xi(\lambda)$ that the lower bound $Q(\lambda)$ of Section~\ref{sec:ipr} requires.
The value of \probR{1} bounds $\xi(\lambda)$ only from above, and \probR{2} relaxes the uniform-rate assumption of Section~\ref{sec:pro-form}, so neither certifies a lower bound.

\subsection{Numerical Performance}\label{sec:solver-bench}

Table~\ref{tab:d1-solvers} compares the per-EV solvers on $8192$ EVs drawn from the case-study fleet of Section~\ref{sec:setup}, with $\underline{\delta} = 0.2$ and $\kappa = 10^{-3}$.
Each EV faces load and port prices of the form \eqref{eq:lse}, with about $30\%$ of the feeders it visits congested, at the two port weights $w_{\rm n} \in \{0, 3\}$ of the case study.
All solvers are implemented in \texttt{Python} on \texttt{NumPy}, and the stage DP is additionally compiled to machine code.
Each solver runs on three random draws of these prices, and gaps are averaged over the draws.
Times are wall clock for the full set on one core of a $2.1$~GHz Intel Xeon Skylake 6130 processor, averaged over the last two draws after a warm-up draw.
They exclude compilation and the price-independent sets $\mathcal{B}_j$, which IPR computes once per run.
The reference is a per-EV Gurobi MIQP at its default $10^{-4}$ relative tolerance, and gaps are signed per-EV differences to the reference objective, relative to $\max\{1, |{\rm obj}^\star|\}$, so a solver may attain a lower objective value than the reference.

In objective value, the compiled stage DP is below the reference on about $8\%$ of the instances, at most machine precision above it on at least $95\%$, and at most $10^{-5}$ above it on $99\%$.
Its larger gaps occur on about $0.1\%$ of the instances, where the piece cap binds, and they raise the average gap above the $99$th percentile.
It solves the full set in $3.5$ to $6.7$ seconds, which is $20$ to $36$ times faster than Gurobi and $29$ to $45$ times faster than its pure-Python implementation.

The relaxation-based solver, labeled dual ascent, relaxes the cumulative bounds of \probSP{} with multipliers, under which the problem separates into one on/off comparison per stay, maps the relaxed solution into $\X$ by a forward sweep through the sets $\mathcal{B}_j$ of \eqref{eq:Bj}, and optionally re-solves \probR{1} for the resulting pattern.
Even with the \probR{1} re-solve, its average gap is $5\%$ at $w_{\rm n}=0$ and $11\%$ at $w_{\rm n}=3$, since the relaxation selects wrong plug-in patterns that the convex re-solve cannot correct, and its batched runtime of $5.5$ to $6.6$ seconds is at best comparable to the compiled DP.

Under bidirectional charging, the compiled DP remains about six times faster than Gurobi (Appendix~\ref{appx:v2g}).
Since the certificate of Section~\ref{sec:ipr} requires exact per-EV values, the compiled DP is the per-EV solver in every run of Section~\ref{sec:res}.

\begin{table}[htbp]
\centering
\caption{\textbf{Per-EV best-response solver benchmark.}}
\label{tab:d1-solvers}
\fitcolumn{%
\begin{tabular}{@{}lrrrr@{}}
\toprule
 & time (s)\textsuperscript{a} & \multicolumn{3}{c}{gap to Gurobi}\\
\cmidrule(l){3-5}
 & & avg & p95 & p99\\
\midrule
\multicolumn{5}{@{}l}{$w_{\rm n} = 0$}\\
Gurobi (reference)            & 135.4       & --                  & --                  & --\\
stage-DP, Python              & 195.4       & $1.4\times10^{-5}$  & $2\times10^{-16}$   & $1.3\times10^{-9}$\\
stage-DP, compiled            & \textbf{6.7}& $2.7\times10^{-5}$  & $4\times10^{-16}$   & $6.4\times10^{-6}$\\
dual ascent, pattern re-solve & 90.6 / 6.4  & $5.1\times10^{-2}$  & $1.1\times10^{-1}$  & $1.0$\\
dual ascent, projection only  & 65.0 / 5.5  & $5.9\times10^{-2}$  & $1.3\times10^{-1}$  & $1.0$\\
\midrule
\multicolumn{5}{@{}l}{$w_{\rm n} = 3$}\\
Gurobi (reference)            & 126.2       & --                  & --                  & --\\
stage-DP, Python              & 158.6       & $1.4\times10^{-5}$  & $3\times10^{-16}$   & $7\times10^{-12}$\\
stage-DP, compiled            & \textbf{3.5}& $2.0\times10^{-5}$  & $4\times10^{-16}$   & $2.2\times10^{-8}$\\
dual ascent, pattern re-solve & 95.2 / 6.6  & $1.1\times10^{-1}$  & $5.4\times10^{-1}$  & $1.9$\\
dual ascent, projection only  & 66.0 / 5.5  & $1.3\times10^{-1}$  & $5.8\times10^{-1}$  & $1.9$\\
\bottomrule
\end{tabular}%
}
\par\smallskip
\begin{minipage}{\columnwidth}
\footnotesize
 $8192$ EVs, $\underline{\delta} = 0.2$, single core. Total solve time and signed relative optimality gap to the Gurobi reference.\par
\textsuperscript{a}\,Dual-ascent cells present both the serial and the batched ($B{=}1024$) time.\par
\end{minipage}
\end{table}

\section{San Francisco Bay Area Case Study}\label{sec:res}

\subsection{Setup}\label{sec:setup}

The case study covers the San Francisco Bay Area at $30\%$ EV adoption: $1.80$ million EVs with week-long itineraries synthesized from regional mobility data~\cite{ju2026scalable}, and $1{,}303$ distribution feeders with time-varying hosting capacities published by the utility~\cite{pge_ica_map}.
Each feeder $s$ also has a port capacity $C^{\rm n}_s$, the number of public level-2 ports in its service region, counted from the charging-station inventory of the Alternative Fuels Data Center (AFDC)~\cite{us_department_of_energy_2024}.
These $17{,}336$ ports, as listed in June 2026, are concentrated: $78$ feeders hold half of them, and $556$ hold none.
Drivers in single-family dwellings, $75\%$ of the fleet, are assumed to own a private home charger, whose home stays occupy no shared port (Section~\ref{sec:pro-form}), and all other drivers charge only at public ports, home stays included.
An EV may plug in at any in-region stay and charge at up to $10$~kW, but its public plug-ins count against the port capacity of the feeder, and any excess is a port violation.
We model DC fast charging only for the residual energy demand that level-2 charging at the destinations cannot meet, and we do not coordinate it.
This residual amounts to $3.8\%$ of the weekly energy demand and occurs on $48{,}905$ EVs.
We refer to \textsf{MAC}~\cite{ju2026scalable} for further details of the data processing, and Fig.~\ref{fig:data-maps} in Appendix~\ref{appx:figs} maps the instance.

Every EV has an $80$~kWh battery and faces the plug-in floor $\underline{\delta} = 20\%$, so each plug-in must charge at least $16$~kWh.\footnote{To restore feasibility, the preprocessing of Remark~\ref{rmk:curtail-model} curtails $0.2\%$ of the driving energy, on $0.8\%$ of the EVs.}
The disutility constant is $\kappa = 10^{-3}$.
Appendix~\ref{appx:sensitivity} tests the results against three modeling choices: the plug-in floor, the charger conversion efficiency, and the balancing horizon over which each EV restores its stored energy.

\begin{figure}[htbp]
    \centering
    \includegraphics[width=\columnwidth]{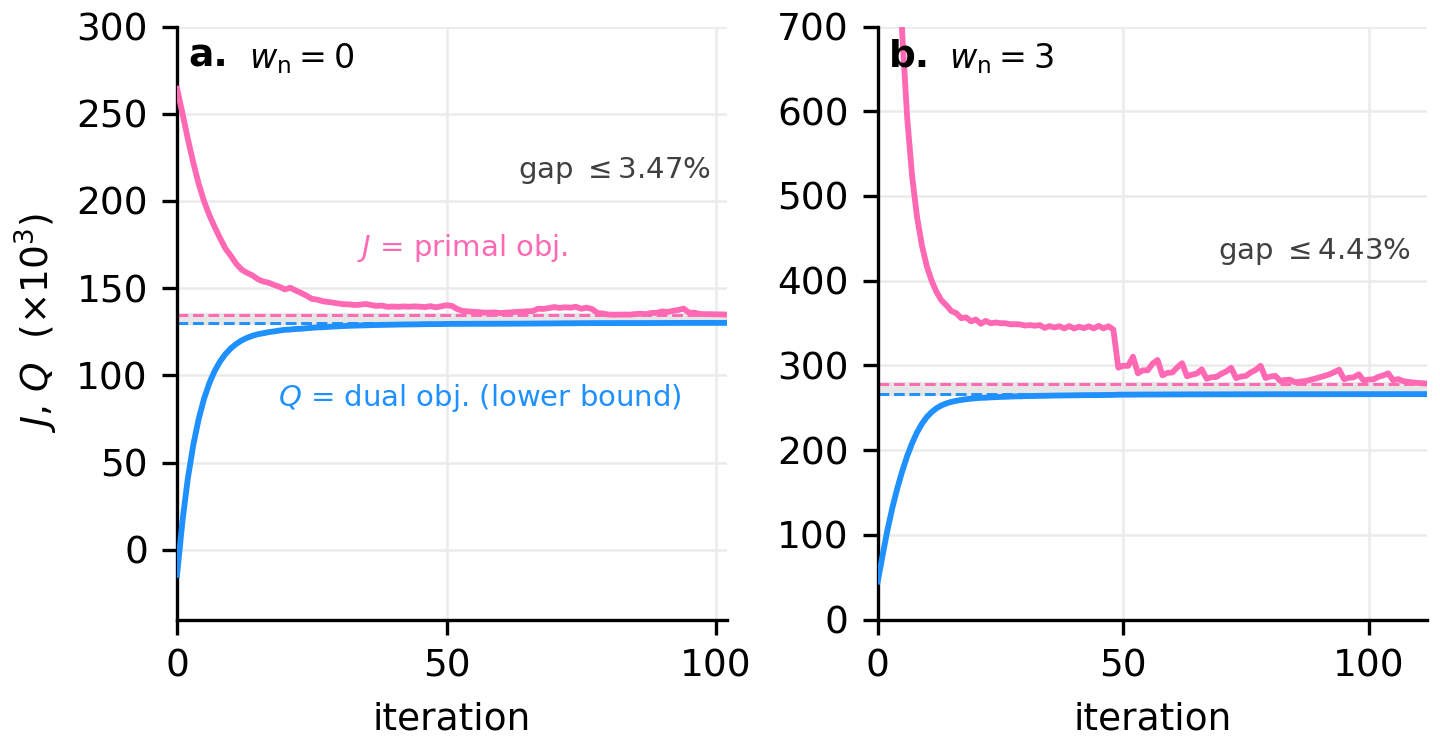}
    \caption{\textbf{Convergence and certificates of IPR}
    \scriptsize\quad
    Deployed objective $J$ and dual bound $Q$ per iteration at
    \textbf{a.} $w_{\rm n} = 0$ and \textbf{b.} $w_{\rm n} = 3$, both
    including the disutility $\sum_i \Jev_i$. Dashed lines mark the best
    pair, and the band between them is the certified optimality gap.
    }
    \label{fig:ipr-convergence}
\end{figure}

Each experiment runs on $10$ nodes of Berkeley's Savio cluster, with two 16-core Intel Xeon Skylake 6130 processors and $96$~GB of memory per node.
Each fleet sweep solves the pricing problems of all EVs with the compiled stage DP of Section~\ref{sec:s1-dp}, distributed by \texttt{Dask} over $80$ single-core workers.
Each IPR run starts from the shortage seed of Algorithm~\ref{alg:ipr}, performs at most $200$ tariff updates, and reports its best certified pair, so every \textsf{MAC+} number below is the response of the fleet to a posted tariff, with a certified gap.
Fig.~\ref{fig:ipr-convergence} traces the two bounds at $w_{\rm n} = 0$ and $3$.
Both bounds improve fastest over the first twenty updates and then level off, with certified gaps of $3.47\%$ and $4.43\%$, respectively.

\subsection{Baselines and the Coordination Spectrum}\label{sec:baselines}

We compare \textsf{MAC+} with a spectrum of baseline policies~\cite{ju2026scalable} that differ in two respects.
The first is the flexibility model. A \emph{session-based} policy fixes the energy of a session with a rule applied at that stay, while a \emph{mobility-aware} policy can move energy between any two stays of the itinerary.
The second is how far a policy coordinates the fleet, from not at all, through a single price signal, to an optimization certified against the global optimum.

\textsf{ASAP+} and \textsf{ASAN} charge at full power from arrival and differ in how much.
At every stay it reaches below half charge, \textsf{ASAP+} charges until the battery is full or holds what the rest of the week requires, and elsewhere it charges only what feasibility forces.
At each stay, \textsf{ASAN} charges the smallest of three amounts: what fills the battery, what reaches the next stay of three hours or more with half the battery left, and what its end-of-week target still lacks.
Their \textsf{u}-prefixed variants deliver the same session energies at a constant rate over each stay.
\textsf{FLEX} keeps the session energies of either rule and optimizes their timing within each stay against the system objective, the best that coordination can do with session-based flexibility.

Among the mobility-aware baselines, \textsf{self-min} lets each EV minimize, on its own, a smooth surrogate of its peak charging power plus $w_{\rm n}$ times its public port-hours, with the peak itself at $w_{\rm n} = 0$.
It is uncoordinated, since each EV accounts for its own port use but not for where ports are scarce.
\textsf{PR} posts the price \eqref{eq:lse} computed from the load and port shortages of \textsf{ASAP+} and lets the fleet respond once, and \textsf{MAC+} is IPR run to its best certified pair.

We score every policy by the two terms of \eqref{eq:Jsys}: the peak hosting-capacity violation $\Jgrid$, labeled ``HC'' and reported in MW, and the peak port-capacity violation $\Jport$, labeled ``EVSE'' and reported in thousands of ports.
They serve as proxies for the feeder and port upgrades required to implement each policy.
To monetize them into an upgrade investment, we assume unit costs of $\$1{,}000$ per kW of hosting capacity and $\$3{,}000$ per installed port, which yields $w_{\rm n} = 3$.

\begin{table}[htbp]
\centering
\caption{\textbf{Upgrade need of every policy.}}
\label{tab:baselines}
\fitcolumn{%
\begin{tabular}{@{}lrrr@{\hskip 14pt}lrrr@{}}
\toprule
\multicolumn{4}{c}{\textbf{Rules and single responses}} & \multicolumn{4}{c}{\textbf{Optimized counterparts}}\\
\cmidrule(r){1-4}\cmidrule(l){5-8}
 & HC & EVSE & invest. & & HC & EVSE & invest.\\
\textit{unit} & MW & $\times 10^3$ & \$M & & MW & $\times 10^3$ & \$M\\
\midrule
\multicolumn{8}{@{}l}{\textit{Session-based}}\\
\textsf{ASAP+}  & 13,326.1 & 645.3 & 15,261.9 & \textsf{FLEX}       & 1,767.4 & -- & 3,703.2\\
\textsf{uASAP+} & 3,613.4  & --    & 5,549.2  & \textsf{FLEX (V2G)} & 1,564.2 & -- & 3,500.0\\
\textsf{ASAN}   & 13,325.7 & 651.5 & 15,280.2 & \textsf{FLEX}       & 704.8   & -- & 2,659.3\\
\textsf{uASAN}  & 2,031.0  & --    & 3,985.5  & \textsf{FLEX (V2G)} & 543.8   & -- & 2,498.3\\
\midrule
\multicolumn{8}{@{}l}{\textit{Mobility-aware}}\\
\textsf{self-min} & 145.1 & 193.2 & 724.7 & \textsf{MAC+}       & 20.3 & 31.2 & 114.0\\
\textsf{PR}       & 87.4  & 135.5 & 493.8 & \textsf{MAC+ (V2G)} & 14.4 & 31.3 & 108.3\\
\bottomrule
\end{tabular}%
}
\par\smallskip
\begin{minipage}{\columnwidth}
\footnotesize
Peak violations of hosting capacity (HC) and of charging-port capacity (EVSE), and the monetized upgrade investment.
Mobility-aware policies at $w_{\rm n} = 3$.
A dash marks a port violation identical by design to that of \textsf{ASAP+} (first two rows) or \textsf{ASAN} (last two rows).
\end{minipage}
\end{table}

\begin{figure}[htbp]
    \centering
    \includegraphics[width=\columnwidth]{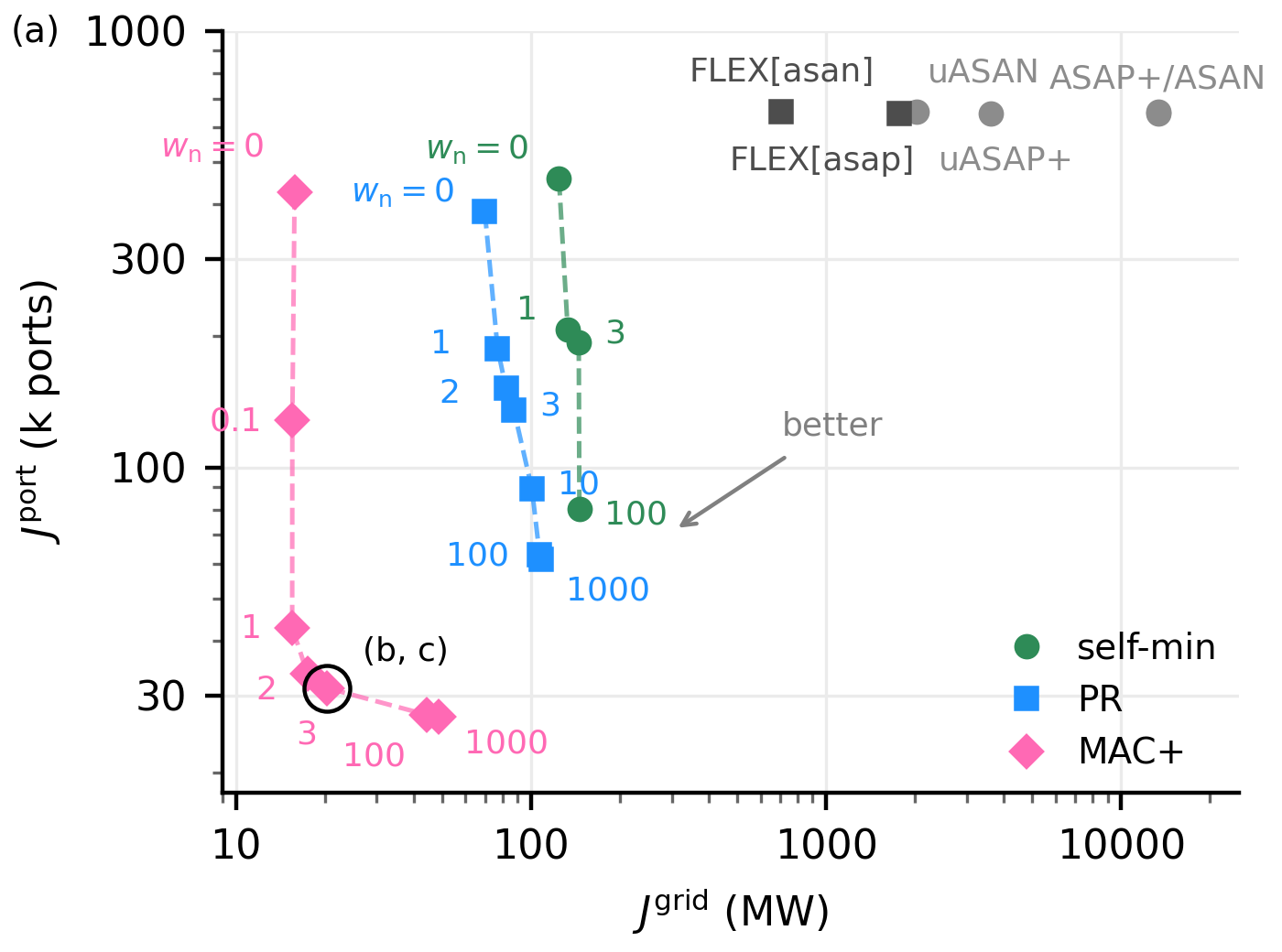}\\[2pt]
    \includegraphics[width=\columnwidth]{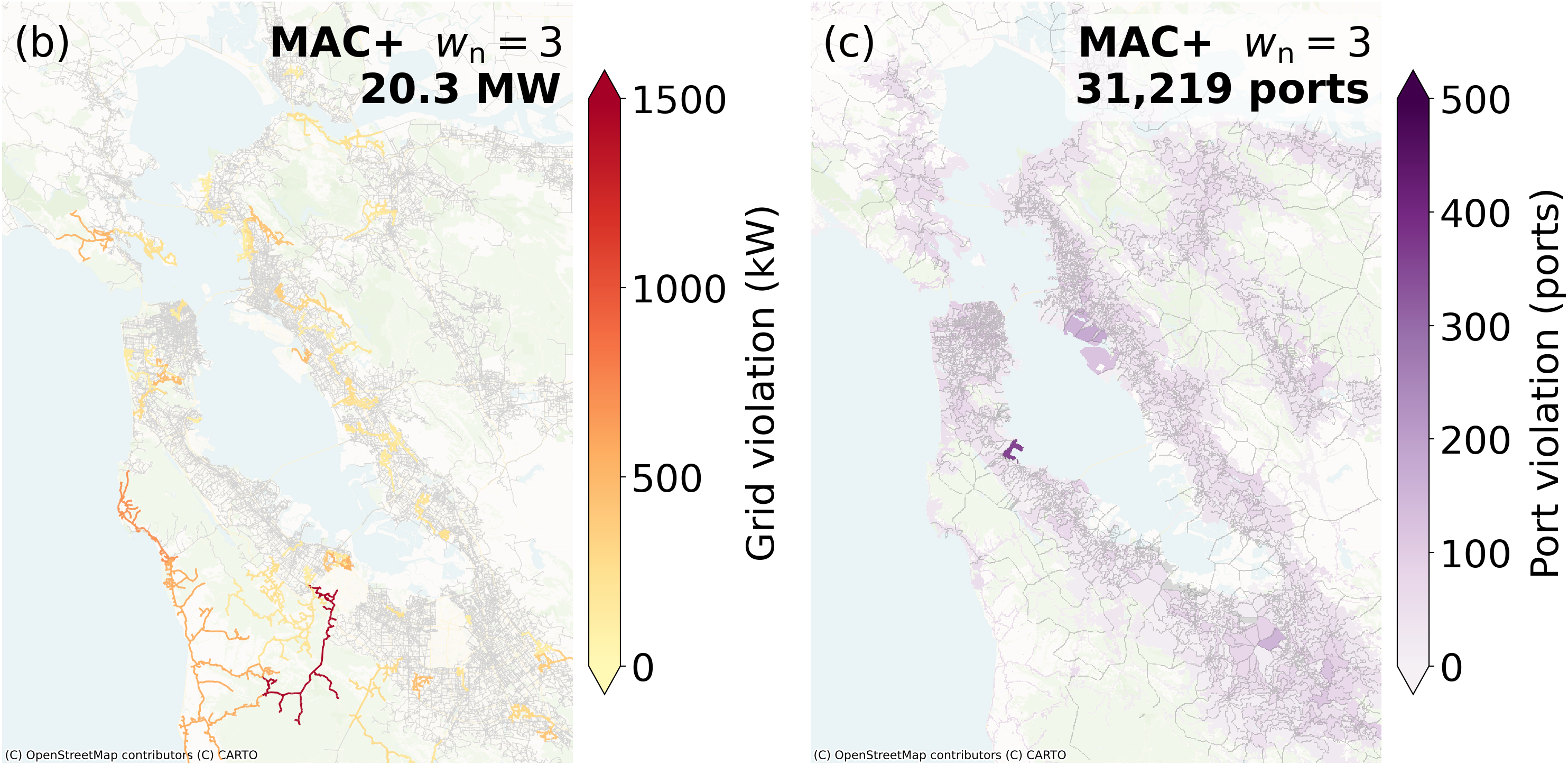}
    \caption{\textbf{The two-capacity frontier and the residual upgrade need}
    \scriptsize\quad
    \textbf{a.} Peak hosting-capacity violation against peak port-capacity violation for every policy, on logarithmic axes.
    The three policies with a port weight are drawn as series annotated with $w_{\rm n}$, and the ringed \textsf{MAC+} point at $w_{\rm n}=3$ is the response mapped in \textbf{b} and \textbf{c}.
    \textbf{b.} Peak hosting-capacity violation of each feeder under \textsf{MAC+} at $w_{\rm n} = 3$, drawn on its distribution lines.
    \textbf{c.} Peak port-capacity violation of each feeder under the same response, drawn on the feeder's service area (the developed land closer to its lines than to those of any other feeder).
    }
    \label{fig:frontier}
\end{figure}

Table~\ref{tab:baselines} reports the three metrics of every policy.\footnote{Because of the plug-in floor, the delivered energy differs slightly across policies. The session rules deliver $94.8$~GWh over the week and \textsf{self-min} $96.2$~GWh, since flattening favors many short plug-ins, each lifted to $16$~kWh by the floor. We score every policy at its own delivered energy, without renormalization.}
Under unmanaged charging, \textsf{ASAP+} violates hosting capacity by $13{,}326$~MW and port capacity by $645$ thousand ports, which calls for an upgrade investment of $\$15.3$ billion.
Spreading its sessions over each stay, \textsf{uASAP+}, lowers the grid violation to $3{,}613$~MW, and optimal re-timing, \textsf{FLEX}, further lowers it to $1{,}767$~MW, at an investment of $\$3.7$ billion.
By design, neither changes the port violation, which accounts for $\$1.9$ billion of the investment.
From the \textsf{ASAN} sessions, the same two steps reach $2{,}031$ and $705$~MW, the latter at an upgrade cost of $\$2.7$ billion.

Uncoordinated mobility-aware charging, \textsf{self-min}, lowers both violations, to $145$~MW and $193$ thousand ports, or $\$725$ million.
It improves on the best session-based coordination, \textsf{FLEX} on the \textsf{ASAN} sessions, on both axes, by $560$~MW and $458$ thousand ports, or $\$1.9$ billion.
The single price response \textsf{PR} reaches $87$~MW, $135$ thousand ports, and $\$494$ million, and \textsf{MAC+} reaches $20.3$~MW, $31$ thousand ports, and $\$114$ million, which is $\$15.1$ billion below \textsf{ASAP+} and $\$380$ million below \textsf{PR}.
Mobility-aware flexibility and coordination are complementary.
The first lowers the investment from $\$2.7$ billion under the best session-based coordination to $\$725$ million without any coordination, and coordination lowers it further to $\$114$ million.

Fig.~\ref{fig:frontier}b and c show where the remaining violations of \textsf{MAC+} occur, and the grid and port parts of its investment are distributed differently.
The grid need is concentrated.
Only $175$ of the $1{,}303$ feeders exceed their hosting capacity at all, $24$ of them account for half of the $20.3$~MW, and $61$ exceed it by more than $0.1$~MW.
The port need is small on each feeder but spread over $1{,}265$ feeders, with $25$ ports per feeder on average, and only $9$ feeders lack $100$ ports or more.
Ports account for $\$94$ million of the $\$114$ million, so the \textsf{MAC+} investment is mostly a region-wide addition of public ports, complemented by upgrades of a few dozen feeders.
Fig.~\ref{fig:infra-maps} in Appendix~\ref{appx:figs} draws the same maps for the other policies.

The \textsf{FLEX (V2G)} and \textsf{MAC+ (V2G)} rows let an EV discharge during a stay (Appendix~\ref{appx:v2g}) under the same tariff mechanism.
For \textsf{MAC+}, V2G lowers HC from $20.3$ to $14.4$~MW and the investment by $\$5.7$ million, while the port violation barely moves, since the port price suppresses the extra plug-ins that discharging would otherwise add.
Even if V2G-capable chargers cost no more than uni-directional ones, V2G thus reduces the investment by only about $5\%$.

\subsection{The Port-Weight Frontier}\label{sec:frontier}

The weight $w_{\rm n} = 3$ rests on assumed unit costs, whereas the relative cost of a port and of feeder headroom varies across sites and utilities.
Three policies take $w_{\rm n}$ as a parameter, \textsf{MAC+}, \textsf{PR}, and \textsf{self-min}, and varying it traces their frontiers between the two capacities in Fig.~\ref{fig:frontier}a, with the weights annotated there.

At the positive weights that all three policies share, $w_{\rm n} \in \{1, 3, 100\}$, \textsf{MAC+} improves on both alternatives on both axes, with an investment saving of $58\%$ to $77\%$ over \textsf{PR} and $68\%$ to $84\%$ over \textsf{self-min}.
Without a port price, $w_{\rm n} = 0$, \textsf{MAC+} coordinates only the grid load and leaves a larger port violation than \textsf{PR}, $426.2$ against $385.5$ thousand ports.

At its grid-optimal end, the \textsf{MAC+} frontier reduces the port violation without increasing the grid violation.
Raising $w_{\rm n}$ from $0$ to $1$ reduces the port violation by $90\%$, from $426.2$ to $42.9$ thousand, while the grid violation falls slightly, from $15.8$ to $15.5$~MW.
This is consistent with the plug-in floor.
Consolidating charging into fewer plug-ins removes energy that the floor would otherwise force, which also reduces the feeder load.
A further reduction of the port violation increases the grid violation only beyond $w_{\rm n} = 1$, and at $w_{\rm n} = 100$ the port violation is $27.1$ thousand at $44.1$~MW.
Fig.~\ref{fig:infra-maps-wn} in Appendix~\ref{appx:figs} shows the same trade-off on the map.

Of the two baselines, \textsf{PR} improves on \textsf{self-min} on both axes at every weight, because port scarcity is local, as Fig.~\ref{fig:data-maps} shows, and only a locational price can target it.

\section{Conclusions and Future Work}\label{sec:conclusion}

This paper develops \textsf{MAC+}, a mobility-aware framework for planning the joint expansion of distribution feeders and public charging ports.
It builds on \textsf{MAC}, inheriting its mobility-aware flexibility and its coordination by iterated price response, and extends both to the public ports as a second capacitated resource alongside the feeders.
On the Bay Area case study with 1.8 million EVs, \textsf{MAC+} reaches a certified optimality gap of $4.4\%$ within a few hours.
Each iteration of IPR solves the mixed-integer best responses of all EVs with a custom exact dynamic-programming solver, about 20 times faster than off-the-shelf solvers.

\textsf{MAC+} saves \$15.1 billion of the upgrade investment that unmanaged charging requires, leaving \$114 million for feeder upgrades and new public ports.
It also saves considerably compared with charging that is fully coordinated but flexible only within each session, which requires \$2.7 billion, and with charging that is flexible along the itinerary but uncoordinated, which requires \$725 million.
The case study further examines the co-expansion frontier along the port-to-feeder cost ratio, bidirectional charging, and the sensitivity to the balancing horizon, the charger conversion efficiency, and the plug-in floor.

We see three directions for future work.
First, behavioral models of EV drivers that reflect bounded rationality, partial compliance, and heterogeneous preferences need to be developed and incorporated in place of the exact best response of every EV to the posted tariff.
Second, the coordination needs practical mechanisms for its implementation and for its operation under stochasticity and uncertainty.
Third, joint modeling of power sectors would allow evaluating transmission congestion and carbon emissions.

\renewcommand*{\bibfont}{\footnotesize} %
\printbibliography[title=References]

\bigskip
{\small
\noindent
\textbf{Declaration of LLM usage}\qquad
The authors used LLMs and Coding Agents (primarily Claude Fable 5 in Claude Code and GPT 5.6 Sol) for code implementation and manuscript polish. The authors reviewed and edited the content as needed and take full responsibility for the content.
}

\newrefsection
\newpage

\appendices
\twocolumn[{\makeatletter\def\@captype{table}\makeatother
\begin{minipage}{\textwidth}
\centering
\scriptsize
\setlength{\tabcolsep}{4pt}
\setlength{\aboverulesep}{0.4pt}
\setlength{\belowrulesep}{0.75pt}
\definecolor{macbg}{gray}{0.92}
\definecolor{maclab}{RGB}{0,90,181}
\setulcolor{maclab}
\setuldepth{a}
\DeclareRobustCommand{\macl}[1]{\textcolor{maclab}{\textbf{\ul{#1}}}}
\begin{threeparttable}
\caption{\textbf{Representative large-scale EV--grid integration studies,} compared by focus, by the fidelity of their mobility, power-infrastructure, and charging-infrastructure models, and by methodology.
}
\label{tab:lit}
\newcolumntype{P}[1]{>{\raggedright\arraybackslash}p{#1}}
\begin{tabular}{@{}P{1.00in}P{1.18in}P{0.92in}P{1.05in}P{1.02in}P{1.12in}@{}}
\toprule
\textbf{Study} (scope) & \textbf{Focus} & \textbf{Mobility} & \textbf{Power infras.} & \textbf{Charging infras.} & \textbf{Methodology}\\
\midrule
Muratori '18~\cite{muratori2018impact}\newline US, 200 households, one transformer
 & \macl{Grid stress.} residential peaks and transformer overload from unmanaged charging
 & \macl{Full.} synthetic connected activity chains; year-long, home-anchored
 & \textbf{Aggregate.} residential load plus one transformer capacity as local cap
 & \textbf{Absent.} home charging only, two power levels
 & \textbf{Simulation.} unmanaged charging across adoption levels\\
\midrule
Wolinetz et al.\ '18~\cite{wolinetz2018simulating}\newline BC and Alberta, to 2050
 & \textbf{Cost.} system cost of controlled charging under realistic participation
 & \textbf{Partial.} adoption and use modeled statistically from short diaries
 & \textbf{Generation.} provincial dispatch and capacity evolution
 & \textbf{Absent.} participation endogenous, roughly half enroll
 & \textbf{Aggregated opt.} iterated adoption, use, and system models\\
\midrule
Xu et al.\ '18~\cite{xu2018planning}\newline Bay Area, 6M simulated drivers
 & \macl{Grid stress.} site charging peak; tariff incentives sized
 & \textbf{Partial.} session-based; typical weekday; itinerary bounds session shifts, no SOC
 & \textbf{Absent.} site charging peak as grid proxy
 & \textbf{Station-resolved.} access at observed session sites
 & \textbf{Localized opt.} MILP re-times sessions at one site\\
\midrule
Powell et al.\ '22~\cite{powell2022charging}\newline WECC, 24--48M EVs
 & \macl{Grid stress} + \textbf{Emissions.} peak, ramping, storage need; dispatch-based CO$_2$
 & \textbf{Partial.} session-based, statistically generated; typical days, no itineraries
 & \textbf{Generation.} single-node merit-order dispatch; no networks
 & \textbf{Absent.} home/work access scenarios
 & \textbf{Heuristic control.} timer rules; site-level workplace opt.\ generalized by regression\\
\midrule
Navidi et al.\ '23~\cite{navidi2023coordinating}\newline US, 11 distribution networks, to 2050
 & \macl{Grid stress.} transformer overloads and voltage violations; upgrade needs and peak load
 & \textbf{Partial.} randomly generated EV charging windows; no itineraries
 & \macl{Distribution.} three-phase networks; linearized model in control, power flow for evaluation
 & \textbf{Absent.} home chargers at consumer nodes
 & \textbf{Centralized opt.} perfect-foresight joint control of all DERs per network; benchmark, not implementable\\
\midrule
Wu et al.\ '24~\cite{wu2024planning}\newline Bay Area 2050, $\sim$90\% adoption
 & \macl{Grid stress} + \macl{Charger adequacy.} off-peak shifting; required charger counts
 & \macl{Full.} weekly itineraries; charging tied to individual stays
 & \textbf{Absent.} evening system-peak window as criterion
 & \textbf{Aggregate.} ZIP-level charger power vs.\ demand, post hoc
 & \textbf{Heuristic control.} rule-based session shifting, full compliance assumed\\
\midrule
Li \& Jenn '24~\cite{li2024impact}\newline California, $>$5,000 feeders, 24M EVs by 2045
 & \macl{Grid stress.} feeder overload, monetized to \$6--20B upgrades
 & \textbf{Partial.} trip-based charging events; typical weekday, no SOC
 & \macl{Distribution.} real feeder capacity envelopes; assessment only
 & \textbf{Absent.} location-type shares; sampled session shapes
 & \textbf{Simulation.} bottom-up uncontrolled projection; no coordination\\
\midrule
Li et al.\ '25~\cite{li2025unlocking}\newline Shenzhen, 0.48M PEVs
 & \macl{Grid stress.} city peak shaving; compensation and degradation trade-offs
 & \macl{Full.} individual stay sequences; SOC with future-trip needs
 & \textbf{Aggregate.} one citywide load curve
 & \textbf{Absent.} stay-tied home/work/other access
 & \textbf{Localized opt.} LP per $\sim$3,000-EV batch; no citywide coupling\\
\midrule
Chen et al.\ '25~\cite{chen2025quantifying}\newline Chicago + Bay Area counties
 & \macl{Grid stress} + \macl{Charger adequacy.} virtual-asset value; supportable charging demand
 & \macl{Full.} daily trip chains; per-EV feasible sets, cyclic day
 & \textbf{Aggregate.} county hosting caps; no network
 & \textbf{Aggregate.} access probability from zonal port counts; expansion scales it
 & \textbf{Aggregated opt.} inner approximation of summed feasible sets\\
\midrule
Duan \& Motter '25~\cite{duan2025grid}\newline contiguous US
 & \textbf{Emissions} + \macl{Grid stress.} congestion-induced CO$_2$; sizes targeted line expansion
 & \textbf{Partial.} zonal daily energy totals; no vehicles or trips
 & \textbf{Generation + Transmission.} unit dispatch, DC power flow, three interconnections
 & \textbf{Absent.} any in-zone grid node, daily energy only
 & \textbf{Aggregated opt.} linear programs over zonal EV pools\\
\midrule
Martin et al.\ '25~\cite{martin2025cascading}\newline WECC, up to 2M EVs
 & \textbf{Emissions.} marginal-signal validity at scale
 & \macl{Full.} week-long itineraries; SOC through drives and stays
 & \textbf{Generation.} single-node dispatch, endogenously re-run
 & \textbf{Station-resolved.} per-driver access at visited sites
 & \textbf{Localized opt.} per-EV convex; sequential group cascade re-prices the signal\\
\midrule
Chen et al.\ '25~\cite{chen2025vehicle}\newline US, 432 regions, 15-year life
 & \textbf{Cost.} household bills; GHG via exogenous marginal signals
 & \macl{Full.} synthetic annual itineraries; SOC across days
 & \textbf{Aggregate.} exogenous price/emission signals; no feedback
 & \textbf{Absent.} home L2 assumed; DCFC fallback at a premium
 & \textbf{Localized opt.} per-vehicle rolling-horizon LP, 20-day window\\
\midrule
Hanig et al.\ '25~\cite{hanig2025driving}\newline US, 134 balancing areas
 & \textbf{Emissions.} consequential, with endogenous capacity adaptation
 & \textbf{Partial.} aggregate regional charging profiles
 & \textbf{Generation + Transmission.} endogenous expansion and dispatch
 & \textbf{Absent.} embedded in profile generation
 & \textbf{Aggregated opt.} exogenous EV load into a capacity-expansion program\\
\midrule
Sanvito et al.\ '26~\cite{sanvito2026coordinated}\newline Europe, 2050 system
 & \textbf{Cost} + \macl{Charger adequacy.} system cost; endogenous charging-capacity deployment
 & \textbf{Partial.} aggregate plug-in and availability profiles; no individuals
 & \textbf{Generation + Transmission.} country nodes, capacity expansion; cross-border links only
 & \textbf{Aggregate.} endogenous national charging capacity
 & \textbf{Aggregated opt.} continental capacity-expansion LP; fleets aggregated\\
\midrule
Xu et al.\ '26~\cite{xu2026proactive}\newline Bay Area, 2,198 feeders, 100\% adoption
 & \macl{Grid stress} + \textbf{Cost.} transformer overload; proactive-vs-incremental NPV
 & \textbf{Partial.} session-based; no SOC across sessions
 & \macl{Distribution.} synthetic feeders, power-flow evaluation post hoc
 & \textbf{Absent.} session windows define availability
 & \textbf{Localized opt.} per-session G2V/V2G; grid evaluated afterward\\
\midrule
Li et al.\ '26~\cite{li2026facilitating}\newline Shenzhen, 1.08M PEVs, 4,549 units
 & \macl{Grid stress} + \textbf{Emissions.} generation--demand mismatch; avoided curtailment and shortage
 & \macl{Full.} individual stay sequences mapped to zones
 & \textbf{Aggregate.} per-community feed-in caps; no network
 & \textbf{Absent.} stay-tied V2G availability
 & \textbf{Localized opt.} independent per-community scheduling of allocated tasks\\
\midrule
Zhou et al.\ '26~\cite{zhou2026decentralized}\newline Singapore 2050, 0.7M EVs
 & \macl{Grid stress.} PV absorption; citywide peak versus line flows
 & \macl{Full.} daily trajectories; typical weekday and weekend
 & \textbf{Transmission.} city network, flows evaluated post hoc
 & \textbf{Absent.} available whenever parked
 & \textbf{Localized opt.} deliberate district decoupling; citywide optimum worsens line flows\\
\midrule
\rowcolor{macbg}
\textsf{MAC}~\cite{ju2026scalable}\newline Bay Area, 1,303 feeders, 1.8M EVs
 & \macl{Grid stress.} peak feeder hosting-capacity violation
 & \macl{Full.} week-long itineraries; SOC across all stays
 & \macl{Distribution.} feeder hosting capacities inside the optimization
 & \textbf{Absent.} L2 assumed at every stay
 & \macl{Certified coordination.} convex; dual price ascent, competitive equilibrium\\
\midrule
\midrule
\rowcolor{macbg}
\textsf{\textbf{MAC+}} (this paper)\newline \textsf{MAC} + 17.3k EVSE
 & \macl{Grid stress} + \macl{Charger adequacy.} feeder and port peak violations; co-expansion investment
 & \macl{Full.}\newline same as \textsf{MAC}
 & \macl{Distribution.}\newline same as \textsf{MAC}
 & \macl{Occupancy.} concurrent plug-ins counted against finite shared ports per feeder
 & \macl{Certified coordination.} \textsf{MAC}'s price ascent inherited; exact custom solver for the mixed-integer per-EV problem\\
\bottomrule
\end{tabular}
Bold labels are standardized tiers within each column, and a \macl{blue} label marks a tier that \textsf{MAC+} (this paper) also reaches. Rows in chronological order.
\end{threeparttable}
\end{minipage}
\vspace{1.5\baselineskip}}]

\section{Comparison of Large-Scale Studies}\label{appx:lit}

Table~\ref{tab:lit} compares the large-scale studies reviewed in Section~\ref{sec:intro} with \textsf{MAC} and \textsf{MAC+}.
The \emph{focus} column records the system aspects that a study can change or size with its own model.
\emph{Mobility} is \emph{partial} when a study sees only isolated charging sessions, day-level energy totals, or aggregate charging profiles, \emph{full} when it follows the connected itinerary of each individual, and \emph{adaptive} when the itinerary itself adjusts to charging.
\emph{Power infrastructure} is labeled by the sector a study models: \emph{absent}, one \emph{aggregate} pool, \emph{generation}, \emph{transmission}, or \emph{distribution}.
\emph{Charging infrastructure} follows the resolution of the charger data a study uses: \emph{absent}, with chargers assumed available wherever an EV parks or at certain locations, \emph{aggregate} zonal stock, \emph{station-resolved} supply, and port \emph{occupancy}, where simultaneous plug-ins are counted against a finite stock.
\emph{Methodology} records how a study computes charging schedules: \emph{simulation}, \emph{heuristic control}, \emph{localized} optimization, \emph{aggregated} optimization, \emph{centralized} optimization, or \emph{certified coordination}.
Localized optimization keeps individual vehicles but drops their coupling, and aggregated optimization keeps the coupling but merges the vehicles into pools.
Centralized optimization keeps both, but only for a small network and with perfect foresight.
Studies that model neither the power nor the charging infrastructure are not included.

\section{Nomenclature}\label{appx:nomenclature}

This appendix lists the symbols that carry a physical or algorithmic meaning beyond a single derivation.
Symbols used only within one proof or derivation are omitted.
Such local symbols may reuse a letter from this list, for example the set $A$ in Section~\ref{sec:feas-oracle} and the conversion efficiencies $\eta_{\rm c}$ and $\eta_{\rm d}$ in Appendix~\ref{appx:sensitivity}, but only where the context makes the meaning unambiguous.

\newlength{\nomsymw}\setlength{\nomsymw}{46pt}
\newlength{\nomgap}\setlength{\nomgap}{8pt}
\newcolumntype{Y}{>{\raggedright\arraybackslash}p{\nomsymw}}
\newcolumntype{Z}{>{\raggedright\arraybackslash}p{\dimexpr\columnwidth-\nomsymw-\nomgap\relax}}

\medskip
\noindent\textit{Acronyms}\\[2pt]
{\footnotesize\noindent
\textbf{AFDC}: Alternative Fuels Data Center \quad
\textbf{DP}: dynamic program \quad
\textbf{EV}: electric vehicle \quad
\textbf{EVSE}: electric vehicle supply equipment, here the shared charging ports \quad
\textbf{HC}: hosting capacity \quad
\textbf{IPR}: iterated price response \quad
\textbf{L2}: level-2 charging \quad
\textbf{MAC}: mobility-aware coordinated EV charging, the companion paper \quad
\textbf{MAC+}: mobility-aware co-expansion planning, this paper \quad
\textbf{MIQP}: mixed-integer quadratic program \quad
\textbf{PWQ}: piecewise quadratic \quad
\textbf{SO}: social optimum \quad
\textbf{V2G}: vehicle-to-grid, i.e., bidirectional charging\par}

\medskip
\noindent\textit{Operators}\\*[2pt]
{\footnotesize
\begin{tabular}{@{}Y@{\hspace{\nomgap}}Z@{}}
\toprule
Symbol & Description \\
\midrule
$[\![n]\!]$ & Index set $\{1, \dots, n\}$ \\
$[x]^+$ & Positive part $\max\{x, 0\}$, element-wise for vectors \\
$\ind{\cdot}$ & Indicator, $1$ if the condition holds and $0$ otherwise \\
$\odot$ & Element-wise product \\
$\preceq$, $\succeq$ & Element-wise inequalities \\
$\langle \cdot, \cdot \rangle$ & Inner product, the sum of element-wise products \\
$\|\cdot\|_{s,\infty}$ & Largest absolute entry over the slots of feeder $s$ \\
$\cdot \oplus [a, b]$, $\cdot \ominus [a, b]$ & Shift of a set in $\R$ by every point of $[a, b]$, or of $[-b, -a]$ \\
$\Pi_\Lambda$ & Euclidean projection onto $\Lambda$ \\
\bottomrule
\end{tabular}
}

\medskip
\noindent\textit{Sets and Indices}\\*[2pt]
{\footnotesize
\begin{tabular}{@{}Y@{\hspace{\nomgap}}Z@{}}
\toprule
Symbol & Description \\
\midrule
$i \in \Iset$ & Index of EVs, with $I$ the number of EVs \\
$j \in \JiSet$ & Index of the stays of EV $i$, with $J_i$ its number of stays \\
$t \in \Tset$ & Index of unit-length time slots, with $T$ the horizon length \\
$s \in \Sset$ & Index of feeders, with $S$ the number of feeders \\
$r \in \{{\rm l}, {\rm n}\}$ & Resource, the feeder load (l) or the port occupancy (n) \\
\bottomrule
\end{tabular}
}

\medskip
\noindent\textit{Parameters}\\*[2pt]
{\footnotesize
\begin{tabular}{@{}Y@{\hspace{\nomgap}}Z@{}}
\toprule
Symbol & Description \\
\midrule
$\taij$, $\tdij$ & Arrival and departure times of stay $j$ of EV $i$ \\
$\tau_{i,j}$ & Duration $\tdij - \taij$ of stay $j$ \\
$s_{i,j}$ & Feeder serving the location of stay $j$ \\
$A_i$ & Stay-to-time map in $\R^{T \times J_i}$, with $A_i[t,j]$ the fraction of slot $t$ covered by stay $j$ \\
$\overline{A}_i$ & Column-normalized stay-to-time map, so that $\overline{A}_i\, q_i$ is the slot-level charging profile \\
$M^{\rm l}_i$, $M^{\rm n}_i$ & Aggregation maps in $\R^{ST \times J_i}$ that route the charging energy and the port occupancy of each stay to its feeder \\
$M_i$ & Stacked aggregation map $\mathrm{diag}(M^{\rm l}_i, M^{\rm n}_i)$ \\
$D_{i,j}$ & Driving energy of the trip after stay $j$ \\
$E_{i,0}$ & Initial stored energy of EV $i$ \\
$E^{\max}_i$ & Battery capacity of EV $i$ \\
$\underline{E}^{\rm dep}_{i,j}$ & Optional energy target at departure from stay $j$ \\
$\underline{E}^{\rm arr}_{i,j}$ & Safety threshold on the energy at arrival at stay $j$ \\
$P^{\rm L2}$ & Level-2 charging power \\
$\overline{Q}_{i,j}$ & Charger power cap on the energy of stay $j$, $P^{\rm L2}\, \tau_{i,j}$ \\
$\underline{\delta}$ & Plug-in floor as a fraction of the battery capacity \\
$\underline{Q}_{i,j}$ & Plug-in floor on the energy of stay $j$, $\underline{\delta}\, E^{\max}_i$ \\
$\underline{S}_{i,j}$, $\overline{S}_{i,j}$ & Lower and upper bounds on the cumulative charge after stay $j$ \\
$C^{\rm l}_{s,t}$ & Hosting capacity of feeder $s$ in slot $t$ \\
$C^{\rm n}_{s}$ & Port capacity of feeder $s$, the number of shared ports in its service region \\
$C$ & Capacities $(C^{\rm l}, C^{\rm n})$, with $C^{\rm n}_{s,t} \equiv C^{\rm n}_s$ \\
$w_{\rm n}$ & Port weight, the cost ratio between one shared port and one unit of hosting capacity, with the load weight normalized to $1$ \\
$\kappa$ & Disutility constant \\
\bottomrule
\end{tabular}
}

\medskip
\noindent\textit{Decision Variables and Aggregates}\\*[2pt]
{\footnotesize
\begin{tabular}{@{}Y@{\hspace{\nomgap}}Z@{}}
\toprule
Symbol & Description \\
\midrule
$q_i \in \R^{J_i}_{\ge 0}$ & Charging energies of the stays of EV $i$ \\
$\chi_i \in \{0,1\}^{J_i}$ & Plug-in indicators of EV $i$, $\chi_{i,j} = \ind{q_{i,j} > 0}$ \\
$x_i$, $x$ & Decision of EV $i$, $x_i = (q_i, \chi_i)$, and of the fleet, $x = (x_i)_i$ \\
$e_{i,j}$ & Stored energy of EV $i$ at the end of stay $j$ \\
$\X_i$ & Feasible set of $x_i$, compact and non-convex \\
$\X^{\rm cont}_i$ & Continuous relaxation of $\X_i$ \\
$l_{s,t}$ & Charging load on feeder $s$ in slot $t$ \\
$n_{s,t}$ & Occupancy, the number of occupied shared ports of feeder $s$ in slot $t$ \\
$y$ & Aggregates $(l, n) = \sum_i M_i x_i$ \\
\bottomrule
\end{tabular}
}

\medskip
\noindent\textit{Objectives and Problems}\\*[2pt]
{\footnotesize
\begin{tabular}{@{}Y@{\hspace{\nomgap}}Z@{}}
\toprule
Symbol & Description \\
\midrule
$\Jgrid(l)$ & Peak hosting-capacity violation, $\sum_s \max_t [l_{s,t} - C^{\rm l}_{s,t}]^+$, labeled HC in the results \\
$\Jport(n)$ & Peak port-capacity violation, $\sum_s \max_t [n_{s,t} - C^{\rm n}_{s}]^+$, labeled EVSE in the results \\
$\Jsys(l, n)$ & System penalty $\Jgrid(l) + w_{\rm n}\, \Jport(n)$ \\
$\Jev_i(q_i)$ & Disutility of EV $i$, $\frac{\kappa}{2} \|\overline{A}_i\, q_i\|_2^2$ \\
$J(x)$ & Objective of \probSO{} at $x$ and its aggregates $y = \sum_i M_i x_i$ \\
$J^\star$ & Optimal value of \probSO{} \\
\probSO{} & Social optimum of coordinated charging \\
\probDone{}$_i$ & Pricing problem of EV $i$ \\
\probSP{}, \probR{0} & Pricing problem of one EV with its index dropped (Section~\ref{sec:solver}) \\
\probR{1}, \probR{2} & Restricted pricing problems with a fixed plug-in pattern and with fixed stay energies \\
\bottomrule
\end{tabular}
}

\medskip
\noindent\textit{Prices, Dual Quantities, and IPR}\\*[2pt]
{\footnotesize
\begin{tabular}{@{}Y@{\hspace{\nomgap}}Z@{}}
\toprule
Symbol & Description \\
\midrule
$\lambda^{\rm l}_{s,t}$ & Load price per unit of energy on feeder $s$ in slot $t$ \\
$\lambda^{\rm n}_{s,t}$ & Port price per occupied port-slot on feeder $s$ in slot $t$ \\
$\lambda$ & Two-tier tariff $(\lambda^{\rm l}, \lambda^{\rm n})$, the multipliers of the coupling constraints \\
$\Lambda(w)$ & Capped simplex, the nonnegative prices whose sum over the slots of each feeder is at most $w$ \\
$\Lambda$ & Dual-feasible set $\Lambda(1) \times \Lambda(w_{\rm n})$ \\
$\xi_i(\lambda)$ & Optimal value of \probDone{}$_i$ \\
$x_i(\lambda)$ & Best response of EV $i$, an exact solution of \probDone{}$_i$ \\
$y(\lambda)$ & Deployed aggregates $\sum_i M_i x_i(\lambda)$ \\
$g(\lambda)$ & Deployed shortage $y(\lambda) - C$, a supergradient of $Q$ \\
$Q(\lambda)$ & Dual function, a lower bound on $J^\star$ \\
$Q^\star$ & Dual optimum $\max_{\lambda \in \Lambda} Q(\lambda)$ \\
$\eta$, $\eta_0$ & Step size of the price update and its initial value \\
$c_0$ & Floor of the step normalization in the price update \\
$\widehat{\lambda}$ & Softmax seed price of IPR \\
$\beta$ & Sharpness of the softmax seed price \\
$K$ & Iteration budget of IPR \\
$x^{\rm best}$, $\lambda^{\rm best}$ & Best certified pair, the iterates of least $J$ and of largest $Q$ \\
$\epsilon^{\rm rel}$ & Relative tolerance of the stopping rule \\
$d$ & Number of coupling constraints, $2ST$ \\
$\omega_i$ & Range of $\Jev_i$ over $\X_i$ \\
$b_i$ & Weighted footprint of EV $i$, which bounds the change in the penalty that its decision can cause \\
\bottomrule
\end{tabular}
}

\medskip
\noindent\textit{Per-EV Solvers (Section~\ref{sec:solver}, EV index dropped)}\\*[2pt]
{\footnotesize
\begin{tabular}{@{}Y@{\hspace{\nomgap}}Z@{}}
\toprule
Symbol & Description \\
\midrule
$\ell_j$ & Stay load price, the load price averaged over stay $j$ \\
$\pi_j$ & Stay port price, the port price summed over the slots of stay $j$ \\
$c_j$ & Curvature $\kappa\, \alpha_j$ of the cost of stay $j$, with $\alpha_j = \|\overline{A}[\cdot, j]\|_2^2$ \\
$\Phi(q, \chi)$ & Objective of \probSP{} \\
$\mathcal{R}_j$ & Forward reachable set, the cumulative charges after stay $j$ that the stays up to $j$ can reach \\
$\mathcal{B}_j$ & Backward co-reachable set, the cumulative charges after stay $j$ from which the stays after $j$ can be completed \\
$m$ & Largest number of intervals in any $\mathcal{R}_j$ or $\mathcal{B}_j$ \\
$C_j$ & Cumulative charge after stay $j$, the DP state, distinct from the capacities $C$ \\
$V_j(C)$ & Cost-to-go, the optimal value over the stays from $j$ on when the stays before $j$ have charged $C$ \\
$V^{\rm off}_j$, $V^{\rm on}_j$ & Costs of the no-charge and the charge branch of stay $j$ \\
$\theta$ & Arc, a convex quadratic on an interval, stored as a 5-tuple \\
$\Theta_j$ & Arc set of $V_j$ \\
$\mathcal{T}_j$ & Backward map of stay $j$, $\Theta_j = \mathcal{T}_j(\Theta_{j+1})$ \\
$P$ & Largest arc count $\max_j |\Theta_j|$, capped at $64$ by the piece cap \\
\bottomrule
\end{tabular}
}

\section{Proofs for Section~\texorpdfstring{\ref{sec:pro}}{II}}\label{appx:gap}

\emph{Proof of Lemma~\ref{lem:Lambda}.}
The problem separates over feeders, so fix $s$ and write $u, \lambda, C \in \R^T$.
If $\lambda \succeq 0$ and $\sum_t \lambda_t \le w$, then for every $u$, $\langle \lambda, u - C \rangle \le \sum_t \lambda_t [u_t - C_t]^+ \le w \max_t [u_t - C_t]^+$, so the objective is at least $-\langle \lambda, C \rangle$, with equality at $u = C$.
If some $\lambda_t < 0$, letting $u_t \to -\infty$ leaves the penalty bounded while $-\lambda_t u_t \to -\infty$.
If $\sum_t \lambda_t > w$, taking $u = C + \theta \mathbf{1}$ gives the objective $-\langle \lambda, C\rangle + \theta\,(w - \sum_t \lambda_t)$, which tends to $-\infty$ as $\theta \to \infty$.

\emph{Proof of Proposition~\ref{prop:supergrad}.}
Fix $\lambda, \lambda' \in \Lambda$.
Since $x_i(\lambda) \in \X_i$ is feasible for \probDone{}$_i$ at $\lambda'$ and optimal at $\lambda$, $\xi_i(\lambda') \le \Jev_i(q_i(\lambda)) + \langle \lambda', M_i x_i(\lambda) \rangle = \xi_i(\lambda) + \langle \lambda' - \lambda, M_i x_i(\lambda) \rangle$.
Summing over $i$ and subtracting $\langle \lambda', C \rangle = \langle \lambda, C \rangle + \langle \lambda' - \lambda, C \rangle$ gives, by \eqref{eq:Q-Lambda}, $Q(\lambda') \le Q(\lambda) + \langle y(\lambda) - C, \lambda' - \lambda \rangle$.

\emph{Proof of Theorem~\ref{thm:gap}.}
Let $f_i \coloneqq \Jev_i + \mathbb{I}_{\X_i}$, with $\mathbb{I}$ the convex indicator, and let $f_i^{**}$ be its biconjugate, the convex envelope of $f_i$ over ${\rm conv}\, \X_i$.
Since the coupling is linear and $\Jsys$ is convex with the aggregates free, the Lagrangian dual of \probSO{} coincides with that of its convexification, for which strong duality holds, so $Q^\star = J^{\rm CH} \coloneqq \min \{ \sum_i f_i^{**}(\tilde x_i) + \Jsys(\sum_i M_i \tilde x_i) \}$~\cite{aubin1976estimates, udell2016bounding}.
Let $\tilde x$ attain $J^{\rm CH}$.
For each $i$, the point $(M_i \tilde x_i, f_i^{**}(\tilde x_i))$ lies in the convex hull of $G_i \coloneqq \{(M_i x_i, \Jev_i(x_i)) : x_i \in \X_i\} \subset \R^{d+1}$, so $\sum_i (M_i \tilde x_i, f_i^{**}(\tilde x_i)) \in {\rm conv} \sum_i G_i$.
By the Shapley--Folkman lemma~\cite{starr1969quasi}, this point equals $\sum_i g_i$ with $g_i \in {\rm conv}\, G_i$ for all $i$ and $g_i \in G_i$ for all but at most $d + 1$ indices.
For each of the at most $d+1$ remaining indices, replace $g_i$ by one of the points of $G_i$ in its convex combination.
This changes the aggregate component by at most $b_i$ in the norm of \eqref{eq:footprint} and the cost component by at most $\omega_i$.
The result is a point $\hat x \in \prod_i \X_i$ with $\sum_i \Jev_i(\hat x_i) \le \sum_i f_i^{**}(\tilde x_i) + (d+1)\max_i \omega_i$ and, since each feeder's peak penalty is $1$-Lipschitz in the maximum norm over its slots, $\Jsys(\sum_i M_i \hat x_i) \le \Jsys(\sum_i M_i \tilde x_i) + (d+1) \max_i b_i$.
Hence $J^\star \le J(\hat x) \le J^{\rm CH} + (d+1)\max_i(\omega_i + b_i)$.
The second claim follows by dividing by $J^\star$.
The argument is that of~\cite{bertsekas1983optimal} for unit commitment. The duality gap of a separable program is controlled by the non-convexity of a bounded number of its components, independently of how many there are, while the objective grows with the population.

\section{Proofs for Section~\texorpdfstring{\ref{sec:solver}}{III}}\label{appx:proofs-solver}

As in Section~\ref{sec:solver}, we fix one EV and drop its index.
By \eqref{eq:disjunction}, the constraints of \probSP{} on stay $k$ require $q_k \in \{0\} \cup [\underline{Q}_k, \overline{Q}_k]$, with $\chi_k = \ind{q_k > 0}$, and a cumulative charge after stay $k$ in $[\underline{S}_k, \overline{S}_k]$.

\subsection{Exact Feasibility Test}\label{appx:pf-oracle}

This subsection proves Theorem~\ref{thm:oracle}.

(i) We prove the claim on $\mathcal{R}_j$ by induction on $j$.
At $j = 0$ no stay has been decided, the cumulative charge is $0$, and $\mathcal{R}_0 = \{0\}$.
Let the claim hold at $j - 1$.
The constraints of \probSP{} on stay $j$ involve the decisions before it only through their cumulative charge $c'$, which ranges over $\mathcal{R}_{j-1}$ by the induction hypothesis.
Hence $c$ is the cumulative charge after stay $j$ of decisions that satisfy the constraints on the stays up to $j$ if and only if $c = c' + q_j$ with $c' \in \mathcal{R}_{j-1}$, $q_j \in \{0\} \cup [\underline{Q}_j, \overline{Q}_j]$, and $c \in [\underline{S}_j, \overline{S}_j]$.
The choices $q_j = 0$ and $q_j \in [\underline{Q}_j, \overline{Q}_j]$ give the unshifted and the shifted set in \eqref{eq:Rj}, and the last condition gives the intersection.
The claim on $\mathcal{B}_j$ follows by backward induction.
At $j = J$ no stay remains, so every $c \in [\underline{S}_J, \overline{S}_J]$ qualifies, and $\mathcal{B}_J = [\underline{S}_J, \overline{S}_J]$.
Let the claim hold at $j + 1$.
The constraints on the stays after $j$ involve $c$ and $q_{j+1}$ only through the cumulative charge $c + q_{j+1}$ after stay $j + 1$, and by the induction hypothesis, $\mathcal{B}_{j+1}$ is the set of these charges that meet the cumulative bound of stay $j+1$ and from which the stays after $j + 1$ can be completed.
Hence $c \in [\underline{S}_j, \overline{S}_j]$ qualifies if and only if $c + q_{j+1} \in \mathcal{B}_{j+1}$ for some $q_{j+1} \in \{0\} \cup [\underline{Q}_{j+1}, \overline{Q}_{j+1}]$, that is, if and only if $c \in \mathcal{B}_{j+1} \cup (\mathcal{B}_{j+1} \ominus [\underline{Q}_{j+1}, \overline{Q}_{j+1}])$, which is \eqref{eq:Bj}.

(ii) A pair $(q, \chi)$ lies in $\X$ if and only if it satisfies the constraints of \probSP{} on all stays.
By (i) at $j = J$, such a pair exists if and only if $\mathcal{R}_J \neq \emptyset$.
By (i) at $j = 0$, where $[\underline{S}_0, \overline{S}_0] = \R$ and the cumulative charge before the first stay is $0$, it exists if and only if $0 \in \mathcal{B}_0$.

(iii) Each set is stored as a list of disjoint closed intervals sorted by their left endpoints.
Shifting every interval keeps the list sorted, so the union of a set with its shifted copy takes one linear merge of two sorted lists and one sweep that joins overlapping intervals, and the intersection with an interval takes one more sweep.
Each stage therefore costs $\mathcal{O}(m)$, and each recursion costs $\mathcal{O}(J m)$.
\hfill$\blacksquare$

\subsection{Domains and the Recursion}\label{appx:pf-dom}

This subsection proves Proposition~\ref{prop:dom} and the recursion \eqref{eq:dp} for the functions $V_j$ of \eqref{eq:Vj}.
Recall from Theorem~\ref{thm:oracle}(i) that $\mathcal{B}_{j-1}$ is the set of cumulative charges $c \in [\underline{S}_{j-1}, \overline{S}_{j-1}]$ after stay $j - 1$ from which the stays after $j - 1$ can be completed within the constraints of \probSP{}, with $[\underline{S}_0, \overline{S}_0] = \R$.

\emph{Proof of Proposition~\ref{prop:dom}.}
For $j \le J$, $V_j(C) < \infty$ if and only if the constraints of \eqref{eq:Vj} admit a solution, that is, if and only if the stays from $j$ on can be completed from the cumulative charge $C$ after stay $j - 1$.
For $C \in [\underline{S}_{j-1}, \overline{S}_{j-1}]$, this is the defining property of $\mathcal{B}_{j-1}$.
For $j = J + 1$, $\operatorname{dom} V_{J+1} = \R$ and $\mathcal{B}_J = [\underline{S}_J, \overline{S}_J]$.
At $j = 1$, the box $[\underline{S}_0, \overline{S}_0]$ is $\R$, so $\operatorname{dom} V_1 = \mathcal{B}_0$, and $V_1(0) < \infty$ if and only if $\X \neq \emptyset$ by Theorem~\ref{thm:oracle}(ii).
\hfill$\blacksquare$

\emph{Proof of the recursion \eqref{eq:dp}.}
For every $j$ and $C$, the feasible set of \eqref{eq:Vj} is a finite union of polytopes, one for each plug-in pattern $(\chi_k)_{k \ge j}$, and its objective is continuous, so the minimum is attained whenever the feasible set is nonempty.
Fix $j \le J$ and $C$, and separate the decision of stay $j$ from those after it.
If $\chi_j = 0$, then $q_j = 0$, stay $j$ costs nothing, and its cumulative bound requires $C \in [\underline{S}_j, \overline{S}_j]$.
If $\chi_j = 1$, then $q_j \in [\underline{Q}_j, \overline{Q}_j]$, stay $j$ costs $\frac{c_j}{2} q_j^2 + \ell_j q_j + \pi_j$, and its cumulative bound requires $C + q_j \in [\underline{S}_j, \overline{S}_j]$.
In both cases, the costs and constraints of the stays after $j$ are those of \eqref{eq:Vj} at $j + 1$ with the cumulative charge $C + q_j$, so their minimum is $V_{j+1}(C + q_j)$, also at $j = J$, where $V_{J+1} \equiv 0$.
Minimizing first over the stays after $j$ and then over the decision of stay $j$ gives \eqref{eq:dp}, where $\min \emptyset = +\infty$ covers an infeasible branch.
\hfill$\blacksquare$

\subsection{PWQ Invariance and the Backward Map}\label{appx:pwq}

This subsection proves Proposition~\ref{prop:pwq} and constructs the backward map $\mathcal{T}_j$.
We fix one EV and one stay $j$, and simplify notation: write $V \coloneqq V_{j+1}$ for the incoming cost-to-go and drop the stage index from the data, so the stage carries curvature $c > 0$, load price $\ell$, port price $\pi$, charge box $[\underline{Q}, \overline{Q}]$, and sandwich box $[\underline{S}, \overline{S}]$.

\subsubsection{Representation and Base Case}

We index the arcs of $\Theta \coloneqq \Theta_{j+1}$ by $p$, write arc $p$ as $\theta_p = (\underline{\theta}_p, \overline{\theta}_p, \theta_{p,2}, \theta_{p,1}, \theta_{p,0})$ and its quadratic as $\phi_p \coloneqq \phi_{\theta_p}$, so that $V(w) = \phi_p(w)$ for $w \in [\underline{\theta}_p, \overline{\theta}_p]$ by \eqref{eq:Theta}.
The invariant of Proposition~\ref{prop:pwq} is threefold: finitely many arcs, a domain $\bigcup_p [\underline{\theta}_p, \overline{\theta}_p]$ that is a finite union of closed intervals, and \emph{nonnegative curvature on every arc}, $\theta_{p,2} \ge 0$, even though $V$ as a whole is non-convex across arcs.
The base case $\Theta_{J+1} = \{(-\infty, +\infty, 0, 0, 0)\}$, a single flat arc on the whole line, satisfies it.
It remains to show that $\mathcal{T}_j$, which evaluates \eqref{eq:dp}, preserves the invariant.

\subsubsection{The No-Charge Branch}

$V^{\rm off}(C) = V(C)$ on $[\underline{S}, \overline{S}] \cap \operatorname{dom} V$, so each arc keeps its coefficients and has its interval intersected with the box.
Finiteness, the interval-union domain, and nonnegative curvatures are trivially preserved.

\subsubsection{The Charge Branch is PWQ with Convex Arcs}

Substitute the landing state $w = C + q$ for the charge amount $q$.
Then
\begin{equation*}
V^{\rm on}(C) = \min_{w}\ \big\{ F(C, w) : w \in [C + \underline{Q},\, C + \overline{Q}] \cap [\underline{S}, \overline{S}] \cap \operatorname{dom} V \big\},
\end{equation*}
with
\begin{equation*}
F(C, w) = \frac{c}{2}(w - C)^2 + \ell\,(w - C) + \pi + V(w).
\end{equation*}
Decompose the minimization over the arcs of $V$: for each arc $p$,
\begin{equation}\label{eq:hp}
h_p(C) = \min\big\{ F_p(C, w) : \mathrm{lo}_p(C) \le w \le \mathrm{hi}_p(C) \big\},
\end{equation}
where $F_p(C, w) = \tfrac{c}{2}(w - C)^2 + \ell(w - C) + \pi + \phi_p(w)$ and
\begin{align*}
\mathrm{lo}_p(C) &= \max\{C + \underline{Q},\ \underline{S},\ \underline{\theta}_p\},\\
\mathrm{hi}_p(C) &= \min\{C + \overline{Q},\ \overline{S},\ \overline{\theta}_p\},
\end{align*}
defined on $\{C : \mathrm{lo}_p(C) \le \mathrm{hi}_p(C)\}$, and $V^{\rm on} = \min_p h_p$.

\emph{Claim: each $h_p$ is a convex PWQ function of $C$ with finitely many arcs.}
Three observations establish it.
\begin{enumerate}
\item \emph{Joint convexity.}
$F_p$ is jointly convex in $(C, w)$: the proximal term $\tfrac{c}{2}(w - C)^2$ is a convex quadratic composed with a linear map, the term $\ell(w - C)$ is linear, and $\phi_p$ is convex because $\theta_{p,2} \ge 0$ by the inductive invariant.
The constraint set $\{(C, w) : \mathrm{lo}_p(C) \le w \le \mathrm{hi}_p(C)\}$ is an intersection of affine half-planes, hence convex.
Partial minimization of a jointly convex function over a convex set is convex, so $h_p$ is convex on each connected component of its domain; in particular, every quadratic arc it produces has nonnegative curvature.
\item \emph{Affine minimizer per region.}
For fixed $C$, $F_p(C, \cdot)$ is strictly convex in $w$ with curvature $c + \theta_{p,2} \ge c > 0$, so the minimizer is the stationary point clipped to the window,
\begin{align}\label{eq:wstar}
& w(C) = \mathrm{clip}\big( w^{*}(C),\ \mathrm{lo}_p(C),\ \mathrm{hi}_p(C) \big),\\
& w^{*}(C) = \frac{c\,C - \ell - \theta_{p,1}}{c + \theta_{p,2}},\notag
\end{align}
obtained from $\partial_w F_p = c(w - C) + \ell + \theta_{p,2} w + \theta_{p,1} = 0$.
All of $w^{*}$, $\mathrm{lo}_p$, $\mathrm{hi}_p$ are piecewise affine in $C$: $w^{*}$ has slope $c/(c + \theta_{p,2}) \in (0, 1]$, and the window bounds are a maximum and a minimum of three affine functions with slopes $1$ or $0$.
Comparisons among finitely many affine functions change order at finitely many points, so the real line splits into finitely many intervals on each of which the clipped minimizer $w(C)$ equals one fixed affine \emph{selector} from
\begin{equation}\label{eq:selectors}
w(C) \in \big\{ w^{*}(C),\ C + \underline{Q},\ C + \overline{Q},\ \underline{S},\ \overline{S},\ \underline{\theta}_p,\ \overline{\theta}_p \big\}.
\end{equation}
\item \emph{Quadratic value per region.}
On each such interval, $w(C) = \sigma C + \varsigma$ is affine, so $h_p(C) = F_p(C, \sigma C + \varsigma)$ is a quadratic in $C$ whose coefficients follow by expansion.
Its curvature depends only on the selector: $\tfrac{c\,\theta_{p,2}}{c + \theta_{p,2}}$ for the interior selector $w^{*}$ (a Moreau-type smoothing of $\theta_{p,2}$ by $c$), $\theta_{p,2}$ for the slope-one selectors $C + \underline{Q}$ and $C + \overline{Q}$, and $c$ for the constant selectors; all are nonnegative, consistent with observation 1).
\end{enumerate}

Finally, $V^{\rm on} = \min_p h_p$ and $V_j = \min\{V^{\rm off}, V^{\rm on}\}$ are pointwise minima of finitely many quadratic arcs.
On the overlap of any two arc domains, two distinct parabolas cross at most twice, so the lower envelope introduces finitely many additional breakpoints, and every envelope piece is a sub-arc of one parent, inheriting its (nonnegative) curvature.
This proves Proposition~\ref{prop:pwq}. \hfill$\blacksquare$

\subsubsection{Domains and the Co-Reachable Sets}

Proposition~\ref{prop:dom} identifies the domains with the sets of \eqref{eq:Bj}, $\operatorname{dom} V_j \cap [\underline{S}_{j-1}, \overline{S}_{j-1}] = \mathcal{B}_{j-1}$, with $[\underline{S}_0, \overline{S}_0] = \R$.
Every entering state of stage $j$ generated by a feasible prefix lies in $[\underline{S}_{j-1}, \overline{S}_{j-1}]$, so $\mathcal{T}_j$ stores $\Theta_j$ only on $\mathcal{B}_{j-1}$, and every landing state after stay $j$ in the forward pass of Algorithm~\ref{alg:dp} lies in $\mathcal{B}_j$.
This is where the sets of Section~\ref{sec:feas-oracle}, computed once per EV, enter the DP.

\subsubsection{The Backward Map as an Algorithm}

Algorithm~\ref{alg:dpstep} summarizes $\mathcal{T}_j$ as implemented, with the data of stay $j$ and the set $\mathcal{B}_{j-1}$ as fixed inputs.
For the interior selector, the validity region is $\{C : \mathrm{lo}_p(C) \le w^{*}(C) \le \mathrm{hi}_p(C)\}$; a boundary selector is valid where $w^{*}$ falls outside the window on its side and that bound is the active one, all of which are intersections of finitely many affine inequalities in $C$, hence intervals with closed-form endpoints.

\begin{algorithm}
\caption{Backward map $\mathcal{T}_j$, $\Theta_{j+1} \to \Theta_j$}\label{alg:dpstep}
\begin{algorithmic}[1]\small
\Procedure{$\mathcal{T}_j$}{$\Theta_{j+1}$}
\State $\mathcal{E} \gets \emptyset$
\Comment{candidate arcs}
\For{each arc $p$ of $\Theta_{j+1}$}
\Comment{no-charge branch}
\State add $\phi_p$ on $[\underline{\theta}_p, \overline{\theta}_p] \cap [\underline{S}, \overline{S}]$ to $\mathcal{E}$
\EndFor
\For{each arc $p$ of $\Theta_{j+1}$}
\Comment{charge branch}
\State compute $w^{*}(C)$ by \eqref{eq:wstar}
\For{each selector $w(C)$ in \eqref{eq:selectors}}
\State $\Gamma \gets$ validity interval of the selector
\If{$\Gamma \neq \emptyset$}
\State expand $F_p(C, w(C))$ in $C$; add the arc on $\Gamma$ to $\mathcal{E}$
\EndIf
\EndFor
\EndFor
\State $\Theta_j \gets$ \textsc{LowerEnvelope}($\mathcal{E}$)
\Comment{sort breakpoints; sweep; insert pairwise parabola crossings}
\State restrict $\Theta_j$ to $\mathcal{B}_{j-1}$; merge identical adjacent arcs
\State optionally apply the piece cap (Appendix~\ref{appx:cap})
\State \Return $\Theta_j$
\EndProcedure
\end{algorithmic}
\end{algorithm}

With $P$ arcs, the envelope sweep dominates and one evaluation of $\mathcal{T}_j$ costs $\mathcal{O}(P \log P)$, giving $\mathcal{O}(J P \log P)$ for the full backward pass, as stated in Section~\ref{sec:s1-dp}.

The forward step of Algorithm~\ref{alg:dp} applies the same construction at one entering state $C$.
Every arc $p$ of $\Theta_{j+1}$ with $\mathrm{lo}_p(C) \le \mathrm{hi}_p(C)$ yields the minimizing landing state $w(C)$ of \eqref{eq:wstar}, and the arc with the least $F_p(C, w(C))$ gives $V^{\rm on}_j(C)$ and $q^{\rm on} = w(C) - C$.
The no-charge value $V^{\rm off}_j(C)$ is $V_{j+1}(C)$, one evaluation of \eqref{eq:Theta}, for $C \in [\underline{S}_j, \overline{S}_j]$ and $+\infty$ otherwise.
Both sweep the sorted arcs of $\Theta_{j+1}$ once, at cost $\mathcal{O}(P)$ per stay.
An implementation may instead record with each arc of $\Theta_j$ its generating branch and selector, which turns the forward step into one interval lookup per stay.

\subsection{Exactness of the Stage DP}\label{appx:pf-exact}

This subsection proves Theorem~\ref{thm:dp-exact}.

\emph{Backward pass.}
By Appendix~\ref{appx:pf-dom}, the functions $V_j$ of \eqref{eq:Vj} satisfy \eqref{eq:dp}, and by Proposition~\ref{prop:pwq}, $\mathcal{T}_j$ maps the arc set of $V_{j+1}$ to that of $V_j$.
Since $\Theta_{J+1}$ describes $V_{J+1} \equiv 0$, backward induction shows that the backward pass returns the arc set $\Theta_j$ of $V_j$ for every $j$.
Evaluating $\Theta_1$ at $C = 0$ by \eqref{eq:Theta} gives $V_1(0)$, the optimal value of \probSP{}, since \eqref{eq:Vj} at $j = 1$ and $C = C_0 = 0$ is \probSP{}.

\emph{Forward pass.}
If $\X \neq \emptyset$, then $V_1(0) < \infty$ by Proposition~\ref{prop:dom}.
Let $V_j(C_{j-1}) < \infty$ at the entering state $C_{j-1}$ of stay $j$.
If $V^{\rm off}_j(C_{j-1}) \le V^{\rm on}_j(C_{j-1})$, the algorithm sets $(q_j, \chi_j) = (0, 0)$, which meets the constraints of stay $j$ because $V^{\rm off}_j(C_{j-1}) = V_j(C_{j-1})$ is finite.
Otherwise $V_j(C_{j-1}) = V^{\rm on}_j(C_{j-1}) < V^{\rm off}_j(C_{j-1})$, so every optimal solution of \eqref{eq:Vj} at stay $j$ and $C = C_{j-1}$, which exists by attainment, plugs in at stay $j$, and its charge attains the minimum in $V^{\rm on}_j(C_{j-1})$.
Hence a minimizing $q^{\rm on}$ exists, and the algorithm sets $(q_j, \chi_j) = (q^{\rm on}, 1)$, which meets the constraints of stay $j$.
In both cases, with $C_j = C_{j-1} + q_j$, \eqref{eq:dp} gives
\begin{equation*}
V_j(C_{j-1}) = \tfrac{c_j}{2} q_j^2 + \ell_j q_j + \pi_j \chi_j + V_{j+1}(C_j),
\end{equation*}
so $V_{j+1}(C_j) < \infty$.
By induction, the pass reaches stay $J$ with the constraints of \probSP{} on every stay satisfied, and summing the identities over $j$ with $V_{J+1} \equiv 0$ gives $\Phi(q, \chi) = V_1(0)$, the optimal value of \probSP{}.
\hfill$\blacksquare$

\subsection{Certified Piece Cap}\label{appx:cap}

This subsection states the piece cap of Section~\ref{sec:s1-dp} and proves that it keeps the certificate valid.
Write \eqref{eq:dp} as $V_j = \mathcal{T}_j V_{j+1}$, with the backward map of Proposition~\ref{prop:pwq} acting on functions.
A \emph{capped} backward pass starts from $\widehat{V}_{J+1} \equiv 0$ and, after each backward map, merges adjacent arcs into a function $\widehat{V}_j$ that satisfies, for a merge budget $\varepsilon_j \ge 0$,
\begin{equation}\label{eq:cap-merge}
\widehat{V}_j \le \mathcal{T}_j \widehat{V}_{j+1} + \varepsilon_j, \qquad j = J, \dots, 1.
\end{equation}
By \eqref{eq:cap-merge}, $\widehat{V}_j$ is finite wherever $\mathcal{T}_j \widehat{V}_{j+1}$ is, so a merge may extend the domain, for instance across a hole of $\mathcal{B}_{j-1}$, but never shrinks it.
The \emph{restricted} forward pass is that of Algorithm~\ref{alg:dp} on $\widehat{V}$ with every cumulative charge kept in $\mathcal{B}_j$.
At the entering state $\widehat{C}_{j-1}$, the no-charge branch is admissible only if $\widehat{C}_{j-1} \in \mathcal{B}_j$, and the charge branch minimizes only over landing states in $\mathcal{B}_j$.
In the forward step of Appendix~\ref{appx:pwq}, this splits the arcs at the endpoints of $\mathcal{B}_j$ and drops their parts outside $\mathcal{B}_j$.

\begin{proposition}[Certified piece cap]\label{prop:cap}
Let $\widehat{V}_{J+1} \equiv 0$, let $\widehat{V}_J, \dots, \widehat{V}_1$ satisfy \eqref{eq:cap-merge}, and let $(\widehat{q}, \widehat{\chi})$ be recovered by the restricted forward pass.
If $\X \neq \emptyset$, then $(\widehat{q}, \widehat{\chi}) \in \X$ and
\begin{equation}\label{eq:cap-bound}
\widehat{V}_1(0) - \sum\nolimits_{j} \varepsilon_j \;\le\; \min_{(q, \chi) \in \X} \Phi(q, \chi) \;\le\; \Phi(\widehat{q}, \widehat{\chi}).
\end{equation}
\end{proposition}
\begin{proof}
The map $\mathcal{T}_j$ is monotone and commutes with adding a constant, so $\widehat{V}_{j+1} \le V_{j+1} + \sum_{j' > j} \varepsilon_{j'}$ implies $\widehat{V}_j \le \mathcal{T}_j \widehat{V}_{j+1} + \varepsilon_j \le V_j + \sum_{j' \ge j} \varepsilon_{j'}$.
Backward induction from $\widehat{V}_{J+1} = V_{J+1}$ gives this bound for every $j$, and $j = 1$, $C = 0$ give the left inequality by Theorem~\ref{thm:dp-exact}.
The bound also gives $\mathcal{B}_j \subseteq \operatorname{dom} V_{j+1} \subseteq \operatorname{dom} \widehat{V}_{j+1}$, where the first inclusion is Proposition~\ref{prop:dom}.
For the right inequality, $0 \in \mathcal{B}_0$ by Theorem~\ref{thm:oracle}, and by \eqref{eq:Bj} every entering state in $\mathcal{B}_{j-1}$ lies in $\mathcal{B}_j$ or reaches $\mathcal{B}_j$ with some $q_j \in [\underline{Q}_j, \overline{Q}_j]$.
Hence the restricted forward pass never stalls, and since $\mathcal{B}_j \subseteq [\underline{S}_j, \overline{S}_j]$, $(\widehat{q}, \widehat{\chi})$ satisfies every constraint of $\X$.
\end{proof}

The bound \eqref{eq:cap-bound} has two consequences.
First, replacing $\xi_i(\lambda)$ in \eqref{eq:Q-Lambda} by its left side keeps $Q$ a valid lower bound, and the deployed response is $(\widehat{q}, \widehat{\chi})$, so the certificate \eqref{eq:rel-gap} survives the cap.
Second, the capped response is a best response up to a gap that the solver computes itself,
\begin{equation}\label{eq:cap-subopt}
\Phi(\widehat{q}, \widehat{\chi}) - \min_{(q, \chi) \in \X} \Phi(q, \chi) \;\le\; \Phi(\widehat{q}, \widehat{\chi}) - \widehat{V}_1(0) + \sum\nolimits_j \varepsilon_j .
\end{equation}

Our implementation caps every arc set at $64$ arcs.
While the cap binds, it merges the adjacent pair of arcs with the smallest curvature difference into one convex quadratic through the outer endpoints and the junction of the pair, and the compiled DP adds the sup-norm deviation of each merge to $\varepsilon_j$.

\section{The Bidirectional (V2G) Extension}\label{appx:v2g}

Section~\ref{sec:stay-constr} restricts the main text to uni-directional charging.
This appendix states the bidirectional model and traces why the method survives it unchanged.
As in Section~\ref{sec:solver}, we fix one EV and drop its index.

\emph{Model.}
At every stay that carries a plug-in floor, the disjunction \eqref{eq:disjunction} gains a mirrored discharge branch,
\begin{equation}\label{eq:disjunction-v2g}
q_j \in [-\overline{Q}_j, -\underline{Q}_j] \cup \{0\} \cup [\underline{Q}_j, \overline{Q}_j],
\qquad
\chi_j = \ind{q_j \neq 0},
\end{equation}
so a plugged-in EV either charges or discharges, at a uniform rate within the stay.
The mirrored floor is what preserves the port structure: both nonzero branches are bounded away from zero, so the indicator $\chi_j$ remains well defined, occupancy remains linear in $\chi$ as in \eqref{eq:agg-ln}, and the port price $\pi_j$ of \eqref{eq:stay-prices} applies verbatim as the on-cost.
A discharging stay is paid the posted load price and pays the port price, and the disutility \eqref{eq:Jev} is even in $q_j$, so discharging incurs the same quadratic wear.
The penalty \eqref{eq:Jsys} is one-sided, so discharge is valued only through the peak violation it removes, and reverse-flow limits are not modeled.
The sandwich bounds \eqref{eq:S-bounds} are unchanged: under a uniform rate of either sign the stored energy is linear within each stay, so its extremes still sit at stay boundaries and the end-of-stay checks still suffice.

\emph{Machinery.}
Three observations carry the method over.
First, the reachability geometry of Section~\ref{sec:feas-oracle} is preserved.
The recursions \eqref{eq:Rj} and \eqref{eq:Bj} each gain one more copy, shifted by the mirrored window, and finite unions of closed intervals are closed under the extra shift, so Theorem~\ref{thm:oracle} holds with three branches in place of two.
The added branch can only enlarge every $\mathcal{R}_j$ and $\mathcal{B}_j$.
In our data discharge fills holes rather than opening them, and the bidirectional sets carry fewer intervals than their uni-directional counterparts.
Second, the stage DP of Section~\ref{sec:s1-dp} is preserved.
The recursion \eqref{eq:dp} gains a discharge branch $V^{\rm dis}_j$, defined exactly as $V^{\rm on}_j$ with the charge window replaced by $[-\overline{Q}_j, -\underline{Q}_j]$, and the argument of Appendix~\ref{appx:pwq} nowhere uses the sign of that window: its bounds remain affine in $C$ with slopes one or zero, each arc contributes the same finite selector family \eqref{eq:selectors} with mirrored slope-one selectors, and Proposition~\ref{prop:pwq} and Theorem~\ref{thm:dp-exact} extend with the same proofs, while the forward pass of Algorithm~\ref{alg:dp} compares three branches instead of two.
Each backward map processes roughly twice the candidate arcs, an empty discharge window recovers the uni-directional solver, and one implementation serves both models.
Third, the price side of Section~\ref{sec:ipr} never uses $q \succeq 0$.
Weak duality \eqref{eq:weak}, Lemma~\ref{lem:Lambda}, Proposition~\ref{prop:supergrad}, and Theorem~\ref{thm:gap} apply to the enlarged $\X$ exactly as stated, so best responses to the same two-tier tariff remain supergradients and every iterate remains certified.
The deployed effect of the extension is reported in Table~\ref{tab:baselines} and Appendix~\ref{appx:sensitivity}.

\begin{table}[htbp]
\centering
\caption{\textbf{Per-EV best-response solver benchmark under V2G.}}
\label{tab:d1-solvers-v2g}
\fitcolumn{%
\begin{tabular}{@{}lrrrr@{}}
\toprule
 & time (s) & \multicolumn{3}{c}{gap to Gurobi}\\
\cmidrule(l){3-5}
 & & avg & p95 & p99\\
\midrule
Gurobi (reference)            & 206.4        & --                  & --                  & --\\
stage-DP, Python              & 275.7        & $1.3\times10^{-3}$  & $7\times10^{-16}$   & $2.4\times10^{-8}$\\
stage-DP, compiled            & \textbf{33.4}& $-1.2\times10^{-7}$ & $6\times10^{-16}$   & $1.3\times10^{-11}$\\
\bottomrule
\end{tabular}%
}
\par\smallskip
\begin{minipage}{\columnwidth}
\footnotesize
$8192$ EVs, $\underline{\delta} = 0.2$, $w_{\rm n} = 3$, single core, with the price draws of Table~\ref{tab:d1-solvers}. Total solve time and signed relative optimality gap to the Gurobi reference.\par
\end{minipage}
\end{table}

\emph{Benchmark.}
Table~\ref{tab:d1-solvers-v2g} repeats the benchmark of Section~\ref{sec:solver-bench} under V2G at $w_{\rm n} = 3$, on the same EVs and price draws.
Gurobi fails to certify optimality on $378$ instances within its per-EV time limit of $60$ seconds, and the gaps exclude these instances.
The compiled DP accepts a response only if \eqref{eq:cap-subopt} certifies it within a relative $10^{-6}$ of the optimum and otherwise falls back to the per-EV Gurobi solve, which it does on $2412$ of the $3 \times 8192$ solves.
With these fallbacks included, it solves the set in $33$ seconds against $206$ for Gurobi, and its gap to Gurobi stays below $10^{-10}$ on $99\%$ of the instances.
The Python DP has no such fallback, so the few EVs on which its piece cap binds raise its average gap.

\section{Sensitivity Analysis}\label{appx:sensitivity}

Three settings of the case study are modeling choices rather than measurements: the conversion efficiency of the chargers, which the main text sets to one, the plug-in floor $\underline{\delta}$, and the one-week horizon over which each EV restores its stored energy.
This appendix quantifies the exposure of the results to each, at the cost-consistent weight $w_{\rm n} = 3$.
In the first two subsections, the \textsf{MAC+} reference is a longer ascent of the Table~\ref{tab:baselines} run, with HC $20.1$~MW at a certified gap of $3.9\%$.

\subsection{Conversion Efficiency}

Real charging hardware loses roughly $5\%$ per conversion.
The stay model supports this directly: a charge of $q$ battery-kWh draws $q/\eta_{\rm c}$ from the feeder, and a discharge returns $\eta_{\rm d}$ per battery-kWh.
Table~\ref{tab:eta-sens} repeats the spectrum at $\eta_{\rm c} = \eta_{\rm d} = 0.95$, with \textsf{FLEX} omitted, since it re-times grid-side session energy and carries no battery-side loss model.

\begin{table}[htbp]
\centering
\caption{\textbf{Efficiency sensitivity.} The HC violation of each policy at the lossless reference and at $\eta_{\rm c} = \eta_{\rm d} = 0.95$, $w_{\rm n} = 3$.}
\label{tab:eta-sens}
\begin{threeparttable}
\begin{tabular}{@{}lrrr@{}}
\toprule
 & \multicolumn{2}{c}{HC (MW) at $\eta_{\rm c} = \eta_{\rm d} =$} & \\
\cmidrule(lr){2-3}
 & 1 & 0.95 & change\\
\midrule
\textsf{ASAP+}            & 13,326.1 &         & $<1\%$\\
\textsf{ASAN}             & 13,325.7 &         & $<1\%$\\
\textsf{uASAP+}           & 3,613.4  & 3,850   & $+6.6\%$\\
\textsf{uASAN}            & 2,031.0  & 2,200   & $+8.5\%$\\
\textsf{self-min}\tnote{a} & 124.3   & 135.9   & $+9.3\%$\\
\textsf{PR}\tnote{b}      & 107.4    & 116.0   & $+8.0\%$\\
\textsf{MAC+}             & 20.1     & 21.4    & $+6.5\%$\\
\textsf{MAC+ (V2G)}       & 14.4     & 17.2    & $+18.9\%$\\
\bottomrule
\end{tabular}
\begin{tablenotes}
\footnotesize
\item[a] The flattening policy at $w_{\rm n} = 0$.
\item[b] Both members of the pair are seeded from the corrected \textsf{ASAP+} profile, so the pair is like for like, and its baseline therefore differs from the \textsf{PR} row of Table~\ref{tab:baselines}.
\end{tablenotes}
\end{threeparttable}
\end{table}

The primary observation is that no conclusion of Section~\ref{sec:res} changes.
The two columns are nearly proportional: every rung of the coordination ladder moves together, and the HC of \textsf{MAC+} is $18.7\%$ of that of \textsf{PR} lossless and $18.4\%$ at $0.95$, and $16.2\%$ and $15.7\%$ of that of \textsf{self-min}.
The port metric is essentially insensitive, since a conversion loss stretches sessions but rarely flips a plug-in decision: the coordinated responses meter $1$ to $2\%$ fewer occupied ports, and the deployed \textsf{MAC+} investment is unchanged to within $0.1\%$, at $\$112.5$ versus $\$112.4$ million, the small HC rise offset by the slightly lower occupancy.
The size of each HC shift follows how a policy sets its power level.
The front-loaded rules draw the full meter rating either way, so the loss only lengthens their sessions, whereas energy-driven levels, the uniform rules and \textsf{self-min}, scale by about $1/\eta_{\rm c}$, and the priced policies land in the same range.
Bidirectional charging is the one real exception: the round-trip loss prunes the marginal discharge opportunities, so the \textsf{MAC+ (V2G)} response pays $+18.9\%$, although at equal efficiency it still carries a $20\%$ lower HC than its uni-directional counterpart.

\subsection{The Plug-In Floor}

The floor $\underline{\delta}$ encodes plug-in effort, a behavioral constant that is calibrated rather than measured.
Table~\ref{tab:delta-sens} solves \probSO{} at $\underline{\delta} \in \{0.1, 0.2, 0.3\}$, that is, minimum sessions of $8$, $16$, and $24$~kWh, with everything else unchanged.

\begin{table}[htbp]
\centering
\caption{\textbf{Floor sensitivity.} The deployed \textsf{MAC+} solution at three plug-in floors, $w_{\rm n} = 3$, monetized as in Table~\ref{tab:baselines}.}
\label{tab:delta-sens}
\begin{tabular}{@{}llrrr@{}}
\toprule
 & \textit{unit} & \multicolumn{3}{c}{$\underline{\delta}$}\\
\cmidrule(l){3-5}
 & & 0.1 & 0.2 & 0.3\\
\midrule
HC      & MW            & 17.6 & 20.1  & 27.6\\
EVSE    & $\times 10^3$ & 26.1 & 30.8  & 38.6\\
invest. & \$M           & 95.8 & 112.5 & 143.3\\
\bottomrule
\end{tabular}
\end{table}

Three findings.
First, the response is monotone and convex in the floor: a $\pm 0.1$ misspecification moves the deployed HC by $-13\%$ and $+37\%$, and the investment by $-15\%$ and $+27\%$.
The exposure is therefore one-sided, since a true floor above the modeled $0.2$ understates violations, increasingly so, while a lower one is conservative.
Second, the ordering is certified rather than observed: each run's dual lower bound exceeds the objective achieved at the next lower floor, so $J^\star(0.1) < J^\star(0.2) < J^\star(0.3)$ follows from weak duality alone, unaffected by how far any single run converged.
Read as a policy lever, lowering the effort floor from $16$ to $8$~kWh improves the social optimum by at least $6.4\%$, a certified value for purely behavioral interventions, such as automatic authentication or wireless charging, that change no part of the coordination itself.
Third, both violations improve together as the floor drops: the floor is a flexibility knob that shifts the whole frontier, unlike $w_{\rm n}$, which trades along it.
One artifact deserves note.
Delivered energy rises with the floor, from $92.9$ to $101.6$~GWh, a boundary effect of the one-week horizon with fixed initial energy, since floor-forced surplus would carry into the next week rather than add demand.
The violations per delivered GWh remain strongly monotone, so the flexibility loss is real beyond the artifact.

\subsection{The Balancing Horizon}\label{appx:balance}

Every result of Section~\ref{sec:res} lets an EV defer charging across days, provided it restores its initial stored energy by the end of the week.
Whether drivers accept multi-day deferral is a behavioral question, so we measure what the horizon is worth.
A finer balancing horizon partitions the week and requires every EV to hold at least its initial stored energy $E_{i,0}$ at its first departure after each partition boundary.
Table~\ref{tab:balance} re-solves \probSO{} for three partitions, down to daily balance, with everything else as in Table~\ref{tab:baselines}, whose \textsf{MAC+} run is the weekly reference here.

\begin{table}[htbp]
\centering
\caption{\textbf{Balancing-horizon sensitivity.} The deployed \textsf{MAC+} solution at $w_{\rm n} = 3$ under finer balancing horizons, monetized as in Table~\ref{tab:baselines}.}
\label{tab:balance}
\begin{tabular}{@{}llrrrr@{}}
\toprule
 & \textit{unit} & \multicolumn{4}{c}{balancing horizon (days)}\\
\cmidrule(l){3-6}
 & & 7 & $3{+}4$ & $2{+}2{+}3$ & $1{\times}7$\\
\midrule
HC      & MW            & 20.3  & 25.8  & 30.5  & 46.0\\
EVSE    & $\times 10^3$ & 31.2  & 47.4  & 63.2  & 97.2\\
invest. & \$M           & 114.0 & 167.8 & 220.0 & 337.5\\
\bottomrule
\end{tabular}
\end{table}

Shrinking the horizon is costly, and the loss is certified.
Each finer partition's dual bound exceeds the objective achieved at the next coarser one, so the ordering follows from weak duality alone, with per-run gaps between $2.3\%$ and $4.4\%$.
Both violations grow monotonically, and daily balance raises the investment from $\$114.0$ to $\$337.5$ million.
The loss is temporal flexibility rather than added demand.
The feasibility repairs move little, with fast-charged energy rising by $9.4\%$ from weekly to daily balance and curtailment falling, so what disappears is energy banked across days.
Balancing weekly rather than daily thus avoids $\$223.5$ million of upgrades, two thirds of the daily-balance bill.

\section{Supplementary Result Figures}\label{appx:figs}

\begin{figure*}[htbp]
    \centering
    \includegraphics[width=0.8\textwidth]{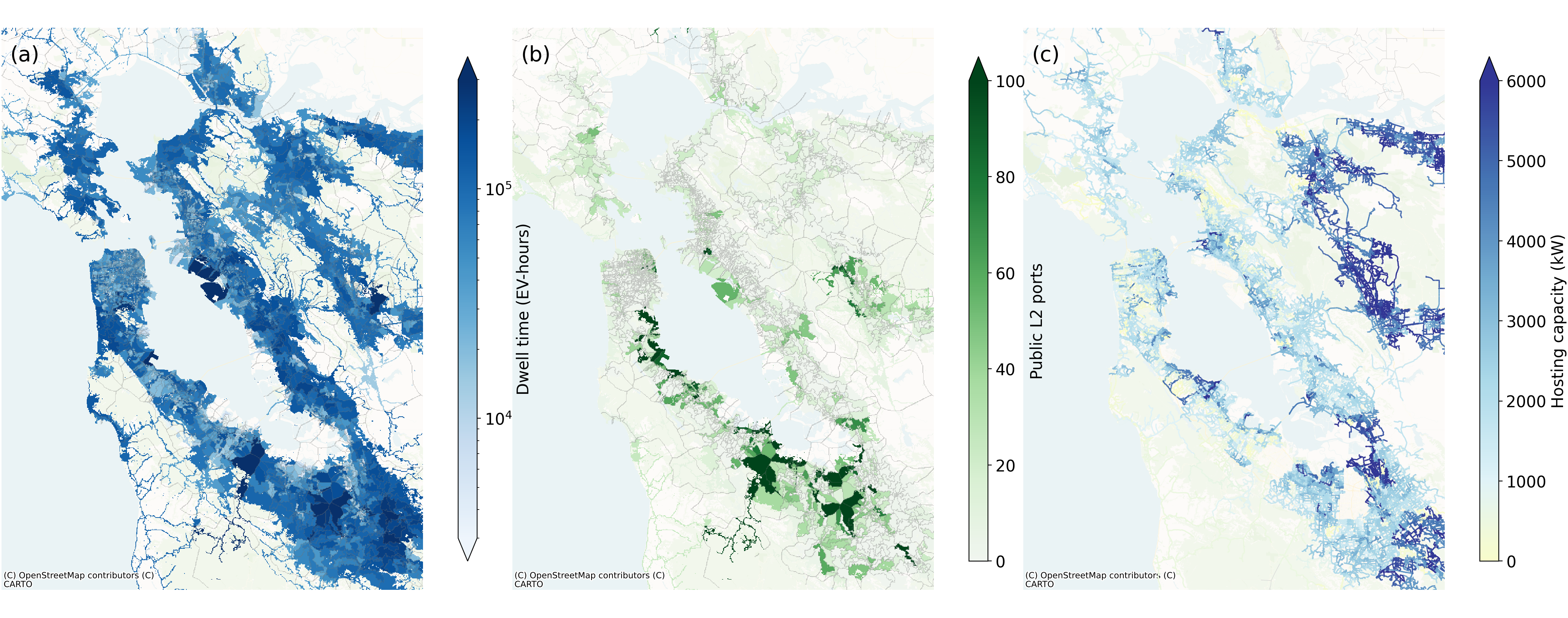}
    \caption{\textbf{The case-study instance}
    \scriptsize\quad
    \textbf{a.} Weekly dwell time of the EV fleet, aggregated per feeder and
    drawn on each feeder's developed service area.
    \textbf{b.} Existing public level-2 charging ports per feeder, on the
    same areas.
    \textbf{c.} Time-averaged hosting capacity, drawn on the feeder lines.
    }
    \label{fig:data-maps}
\end{figure*}

Fig.~\ref{fig:data-maps} maps the case-study instance, whose three layers are poorly aligned.
Dwell time, the opportunity to charge, covers the developed footprint of the region.
The $17{,}336$ existing public ports instead concentrate on few feeders: $78$ feeders hold half of them, and the $556$ feeders with none host $33\%$ of the fleet's dwell time.
The time-averaged hosting capacity ranges from $0.16$ to $6.3$~MW between the 5th and the 95th percentile of the feeders.
Across feeders, the rank correlation of dwell time is $0.26$ with the port count and $0.16$ with the hosting capacity, so where the fleet parks says little about where either capacity lies, and a coordinated response must shift charging across locations as well as across time.

\begin{figure*}[p]
    \centering
    \includegraphics[width=0.9\textwidth]{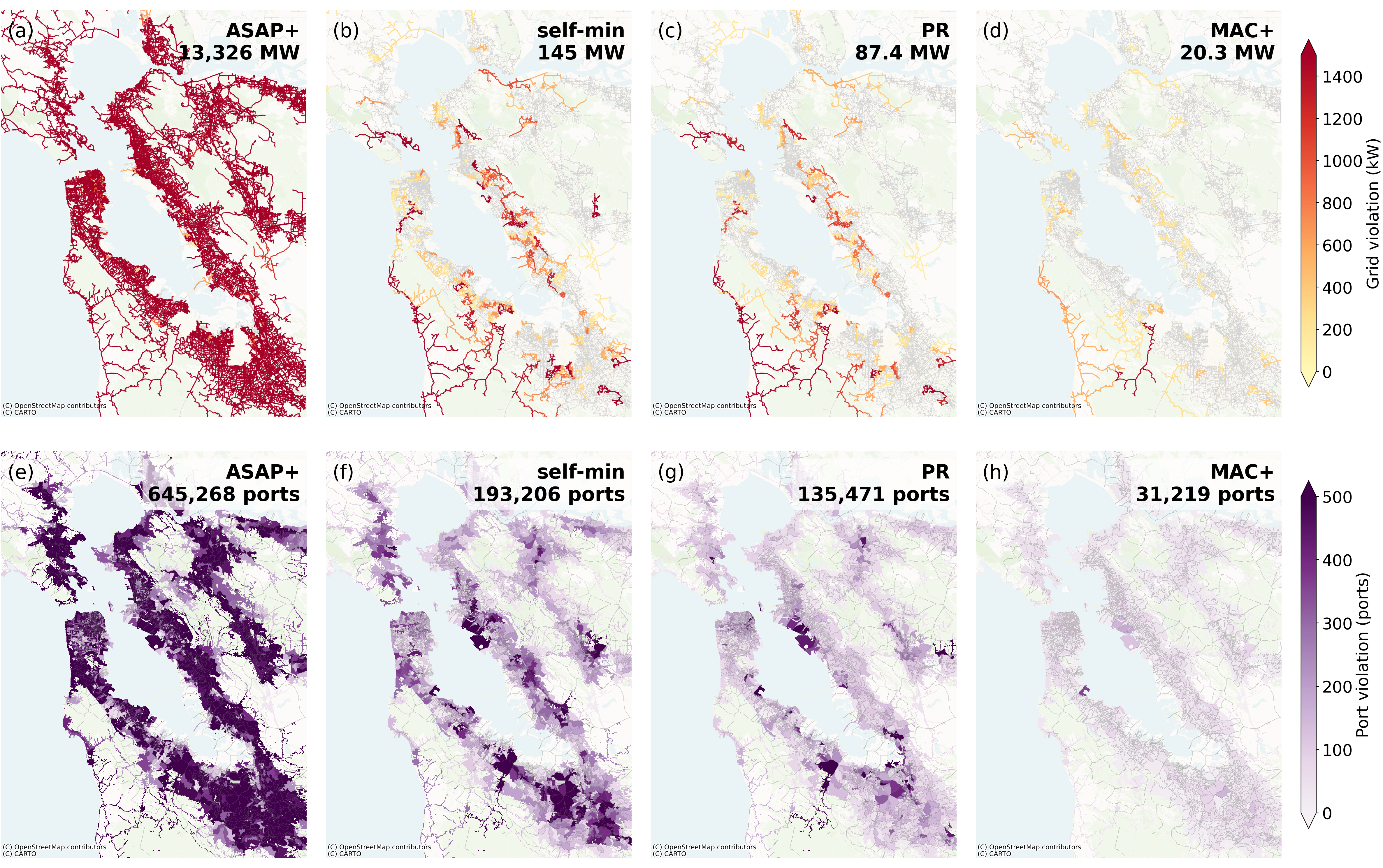}
    \caption{\textbf{Upgrade maps across the coordination spectrum}
    \scriptsize\quad
    Top: peak hosting-capacity violation of each feeder, drawn on its
    distribution lines. Bottom: peak port-capacity violation of each feeder,
    drawn on its service area, the developed land closer to that feeder's
    lines than to those of any other feeder. Color encodes the feeder total,
    shown uniformly across its area. Columns are the policies of
    Table~\ref{tab:baselines}, with \textsf{self-min}, \textsf{PR}, and
    \textsf{MAC+} at $w_{\rm n} = 3$, and the totals over all feeders are
    printed in each panel.
    }
    \label{fig:infra-maps}
\end{figure*}

\begin{figure*}[p]
    \centering
    \includegraphics[width=0.9\textwidth]{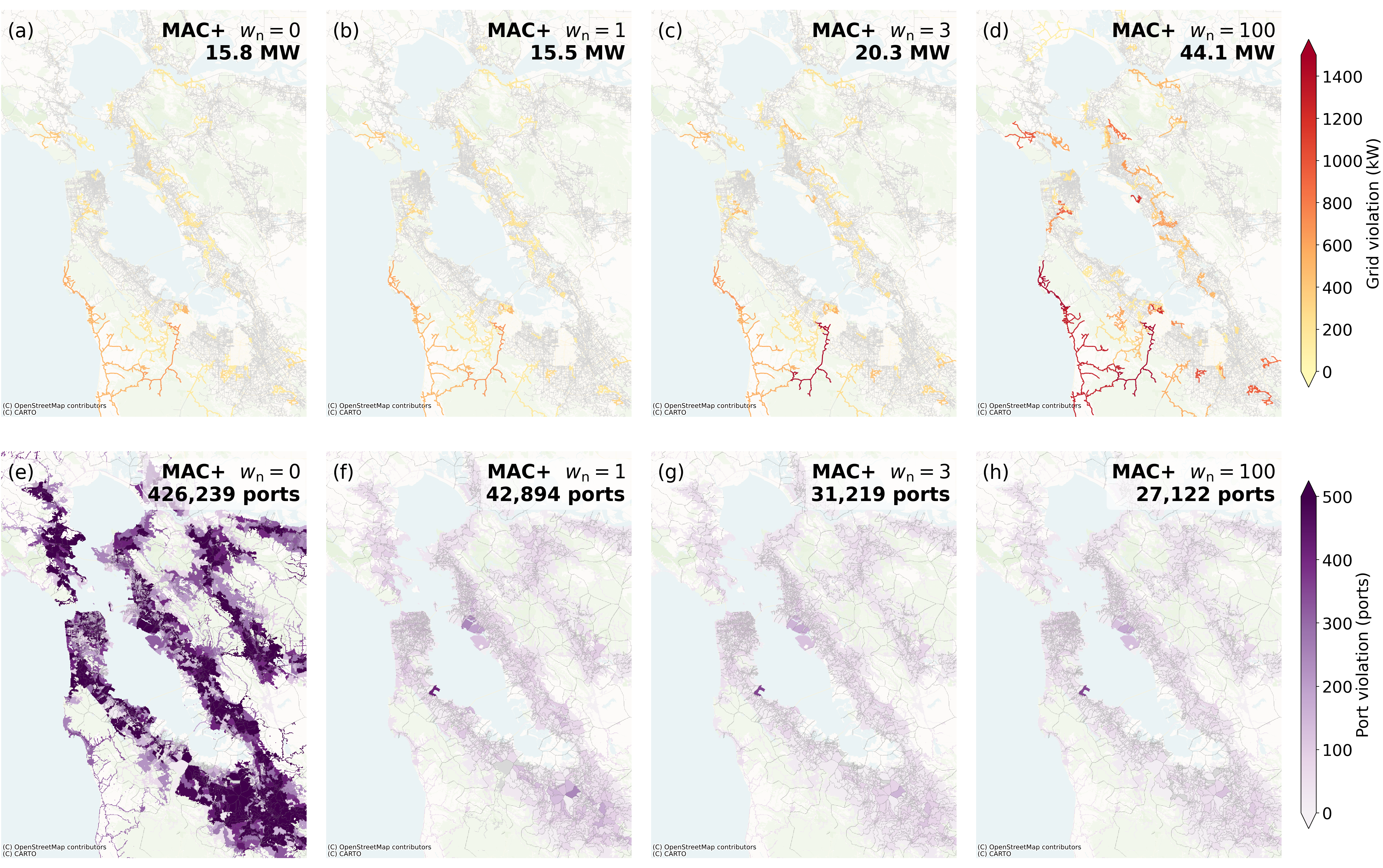}
    \caption{\textbf{Upgrade maps of \textsf{MAC+} across port weights}
    \scriptsize\quad
    The deployed \textsf{MAC+} response at four port weights, drawn as in
    Fig.~\ref{fig:infra-maps}. At $w_{\rm n} = 0$ ports are unpriced, and
    their violation is metered only.
    }
    \label{fig:infra-maps-wn}
\end{figure*}

Fig.~\ref{fig:infra-maps} maps both violations for the policies of Table~\ref{tab:baselines}.
Under \textsf{ASAP+}, $1{,}088$ feeders exceed their hosting capacity by more than $0.1$~MW and $1{,}095$ lack $100$ ports or more, a region-wide build-out on both axes.
Along the spectrum the two violations retreat differently.
The grid violation shrinks in extent, to $203$ feeders above $0.1$~MW under \textsf{self-min}, $150$ under \textsf{PR}, and $61$ under \textsf{MAC+}.
The port violation keeps its extent, with every policy leaving at least one port short on $1{,}262$ to $1{,}284$ feeders, and shrinks in depth instead, as the feeders short of $100$ ports or more number $671$ under \textsf{self-min}, $502$ under \textsf{PR}, and $9$ under \textsf{MAC+}.

Fig.~\ref{fig:infra-maps-wn} maps the \textsf{MAC+} response at four port weights.
At $w_{\rm n} = 0$, the grid optimum leaves $971$ feeders short of $100$ ports or more, so prices that ignore ports leave the port violation in place on the map as well as in aggregate.
The mild weight $w_{\rm n} = 1$ reduces that count to $53$ while the grid map barely changes, with $52$ feeders above $0.1$~MW against $60$ at $w_{\rm n} = 0$.
Larger weights trade grid headroom for the remaining thin port violation.
From $w_{\rm n} = 3$ to $100$, the port violation falls by $13\%$, from $31.2$ to $27.1$ thousand, while the feeders above $0.1$~MW rise from $61$ to $96$.

\clearpage
\printbibliography[title=Supplementary References]

\end{document}